\documentclass[iop,numberedappendix]{emulateapj}

\usepackage{natbib}
\usepackage{amsmath}
\usepackage{graphicx}
\usepackage{upgreek} %for upright greek letters

\usepackage{color, xcolor}
\usepackage[colorlinks=true, linkcolor={magenta},citecolor={blue!50!black},urlcolor={magenta}]{hyperref}

\usepackage{txfonts} 
\usepackage{empheq}

\def\vecbf#1{\mbox{\boldmath $#1$}}

\slugcomment{Accepted for publication in ApJ}
\slugcomment{Draft Version February 12, 2016}

\shorttitle{Light Scattering by fractal dust aggregates}
\shortauthors{Tazaki et al.}

\begin{document}
\title{Light scattering by fractal dust aggregates:\\ I. Angular dependence of scattering}
\author{Ryo Tazaki\altaffilmark{1,2}, Hidekazu Tanaka\altaffilmark{3}, 
Satoshi Okuzumi\altaffilmark{2}, Akimasa Kataoka\altaffilmark{4,5} and Hideko Nomura\altaffilmark{2}
}
\altaffiltext{1}{Department of Astronomy, Graduate School of Science, Kyoto University, Kitashirakawa-Oiwake-cho, Sakyo-ku, Kyoto 606-8502, Japan; \email{rtazaki@kusastro.kyoto-u.ac.jp}}
\altaffiltext{2}{Department of Earth and Planetary Sciences, Tokyo Institute of Technology, 2-12-1 Ookayama, Meguro-ku, Tokyo 152-8551, Japan}
\altaffiltext{3}{Institute of Low Temperature Science, Hokkaido University, Sapporo, Hokkaido 060-0819, Japan}
\altaffiltext{4}{National Astronomical Observatory of Japan, Mitaka, Tokyo 181-8588, Japan}
\altaffiltext{5}{Institute for Theoretical Astrophysics, Heidelberg University, Albert-Ueberle-Strasse 2, 69120 Heidelberg, Germany}

\begin{abstract}
In protoplanetary disks, micron-sized dust grains coagulate to form highly porous dust aggregates.
Because the optical properties of these aggregates are not completely understood, it is important to investigate how porous dust aggregates scatter light. In this study, the light scattering properties of porous dust aggregates were calculated using a rigorous method, the T-matrix method, and the results were then compared with those obtained using the Rayleigh--Gans--Debye (RGD) theory and Mie theory with the effective medium approximation (EMT). The RGD theory is applicable to moderately large aggregates made of nearly transparent monomers.
This study considered two types of porous dust aggregates, ballistic cluster--cluster agglomerates (BCCAs) and ballistic particle--cluster agglomerates (BPCAs). First, the angular dependence of the scattered intensity was shown to reflect the hierarchical structure of dust aggregates; the large-scale structure of the aggregates is responsible for the intensity at small scattering angles, and their small-scale structure determines the intensity at large scattering angles. Second, it was determined that the EMT underestimates the backward scattering intensity by multiple orders of magnitude, especially in BCCAs, because the EMT averages the structure within the size of the aggregates. It was concluded that the RGD theory is a very useful method for calculating the optical properties of BCCAs.
\end{abstract}


\section{Introduction}
%------- General introduction
Protoplanetary disks form around protostars and are thought to be sites of ongoing planet formation.
Planetesimals are formed from dust grains in protoplanetary disks; however, the theory of planetesimal formation involves many problems \citep[e.g.,][]{brauer08, blum08, zsom10}. 
One serious problem is that once dust grains become an approximately meter-sized body, the grains drift radially toward the central star because of the strong head wind of disk gas \citep{whipple72, adachi76, weidenschilling77}. The timescale of this radial drift is much shorter than that of the compact grain growth, and thus the size of the dust grain bodies cannot grow beyond the order of meters. This problem is often referred to as the radial drift barrier. Recent numerical studies have shown that the radial drift barrier and other related problems can be avoided employing the porous aggregation model of icy particles \citep{okuzumi12, kataoka13b}. 
Because of the strong adherence of icy particles to one another due to the hydrogen bonding force, icy dust aggregates are expected not to suffer severe disruption when they experience high-speed collisions \citep{wada09, gundlach11, wada13}. Previous studies have predicted the presence of highly porous dust aggregates in protoplanetary disks, but its presence has not yet been confirmed by observation. 

Aggregates in disks are often modeled using one of two limiting aggregation models, the ballistic cluster--cluster agglomerate (BCCA) and ballistic particle--cluster agglomerate (BPCA) models, which have fractal dimensions of $d_f\lesssim2$ (highly porous) and $d_f\approx3$ (compact), respectively. 
In the initial stage of the coagulation process, the aggregate tends to have $d_f\lesssim2$, as in BCCAs \citep{smirnov90, meakin91, kempf99, blum00}. 
Highly porous dust aggregates are readily stirred up to the surface layer of the disk because of their strong coupling with the disk gas.
Hence, highly porous dust aggregates contribute to the scattering of stellar radiation, 
and their optical properties govern the appearance of the disks at near-infrared wavelengths, including their surface brightness, brightness asymmetry, and color.
As dust aggregates grow larger, they become compressed by mutual collisions, gas compression, and self-gravity \citep{dominik97,  blum00, suyama08, suyama12, paszun09, okuzumi12, kataoka13b}, causing them to settle down to the midplane of the disk. Hence, such compressed aggregates can be observed at millimeter wavelengths.

To investigate how the presence of fractal dust aggregates alters the observed image, their optical properties must be known.
The opacity of dust aggregates has been investigated by various authors \citep[e.g.,][]{kozasa92, kozasa93, henning96, cuzzi14, kataoka14, min06, min08, min15}. More recently, \citet{min15} investigated the light scattering properties of compact dust aggregates. In the present study, the light scattering properties of both fluffy and compact dust aggregates were investigated.

Many previous studies have assumed that dust grains are spherical, which allows the use of the exact Mie solution \citep{Mie1908,bohren83}.
The calculation of the optical properties of nonspherical particles is not an easy task \citep[e.g.,][]{mishchenko00} and is generally achieved by utilizing numerical methods, such as the discrete dipole approximation \citep[DDA;][]{purcell73,draine94} and the T-matrix method \citep[TMM;][]{mishchenko96}. Although these methods can provide correct results, they require long computing times. Hence, it is difficult to use these methods to calculate the radiative transfer of disks with realistic conditions. Although the radiative transfer calculations that take into account the porosity of the grains in the disks have been studied by several authors \citep{min12, Kirchschlager14, murakawa14}, the high computational demands of such methods restricted these studies to the use of a simple dust model or approximations of the optical properties. 
Because of the limitations of radiative transfer calculations, methods of modeling observations are also limited. 
Some near-infrared imaging observations of disks cannot be explained by compact dust grains, and the inconsistencies between the observed properties and the Mie theory results may be attributable to the presence of dust aggregates \citep[e.g.,][]{pinte08, mulders13}. 
These facts motivated us to develop a simple and accurate model of the light scattering properties of dust aggregates. 

Because of its simplicity, the effective medium theory (EMT), or Mie theory using the effective medium approximation, is often used \citep[e.g.,][]{cuzzi14, kataoka14}. 
The EMT allows the effective dielectric function of an inhomogeneous material to be calculated \citep[e.g.,][]{chylek00}.
In the EMT, an aggregate is replaced with a single sphere with a characteristic radius and an effective dielectric function.
Because the derivation of the effective dielectric function usually assumes the presence of electrostatic fields, the EMT is valid only for Rayleigh inclusions, that is, the inclusions are smaller than the incident wavelength. 
When the above condition is satisfied, the EMT tends to yield reliable results for the integrated properties, such as the absorption and scattering opacity \citep[e.g.,][]{bazell90, kozasa92, vosh05,vosh07,min06, shen08}.
However, it fails to reproduce angle-dependent properties, such as the phase function and the polarization, even if the Rayleigh inclusion condition is satisfied \citep{kozasa93, shen09}. 
This is due to the fact that the configuration of particles is ignored, which results in the inability to correctly capture the phase shift of scattered light. % in the EMT calculation.

One method that takes the configuration of a particle into account is the Rayleigh--Gans theory \citep[e.g.,][]{bohren83}. 
The Rayleigh--Gans theory, or the first Born approximation, is applicable to an arbitrarily shaped particle as long as the particle refractive index is close to that of a vacuum, namely, the optically soft. 
In such cases, the total scattered field can be expressed as a combination of the classical Rayleigh scattering solution and the {\it form factor}, which contains information regarding the phase difference arising from the arbitrary shape of the particle.

Furthermore, the Rayleigh--Gans theory can be easily extended to a collection of optically soft particles \citep[e.g.,][]{debye1915}.
The optical properties of arbitrarily shaped dust aggregates can be calculated using the Rayleigh--Gans--Debye (RGD) theory.
In the RGD theory, the scattered field can be described in terms of the classical Rayleigh solution; the form factor of a constituent particle; and the {\it structure factor} of the aggregates, which represents the arrangement of the constituent particles. The RGD theory has been profoundly investigated in the field of atmospheric science \citep[e.g.,][]{sorensen01}.

For astrophysical purposes, \citet{kozasa93} investigated the light scattering properties of fractal dust aggregates using the DDA and compared them with those obtained using the RGD theory. They concluded that the RGD theory is a powerful method of calculating the scattering intensity for a given direction. 
Although the structure factor plays a crucial role in the RGD theory (see Section \ref{sec:analytic} for more detail), the choice of the structure factor is still under debate. 
This paper will clarify how to select a realistic structure factor for BCCA and BPCA models. Using this realistic structure factor, we compare the RGD theory results with the rigorous results obtained using the TMM.
The TMM gives the solution for a collection of spherical particles as a superposition of their exact Mie solutions; this is known to be one of the most rigorous methods of obtaining the optical properties of the aggregates. 
This paper shows that the RGD theory exhibits better agreement with the TMM than the EMT does. Furthermore, the RGD theory captures important qualitative features of the TMM results.

%------- Plan
This paper focuses on the angle-dependent optical properties, such as the phase matrix elements, 
and a subsequent paper in this series will discuss the angle-integrated properties, such as absorption and scattering opacity.
This paper is organized as follows. 
Section \ref{sec:light} summarizes models of light scattering by fractal dust aggregates.
In Section \ref{sec:dust}, the statistical model of fractal dust aggregates, which plays a key role in the RGD theory, is described. Section \ref{sec:results} reports the rigorous results of the phase matrix elements obtained by the TMM and shows that the RGD theory is able to reproduce these rigorous results. The applicability of the RGD theory is discussed in Section \ref{sec:app}. Section \ref{sec:summary} summarizes the results of this study.

\section{Light scattering models} \label{sec:light}
The scattering process can be described by the phase matrix that represents the transition from the vector of the Stokes parameters for the incident light $(I_{\rm inc},Q_{\rm inc},U_{\rm inc},V_{\rm inc})$ to that for the scattered light $(I_{\rm sca},Q_{\rm sca},U_{\rm sca},V_{\rm sca})$.
This paper focuses on the average optical properties of a distribution of randomly orientated dust aggregates.
For example, if the dust aggregates align in the same way owing to some effect, such as a magnetic field, the random orientation assumption would break down. 
However, this paper investigates the situation in which the above condition is naturally satisfied in the disks.
For a distribution of randomly orientated aggregates, the phase matrix can be reduced to
\begin{equation}
   \left(
    \begin{array}{cccc}
	I_{\rm sca} \\
	Q_{\rm sca}\\
	U_{\rm sca}\\
	V_{\rm sca}
    \end{array}
    \right)
   = \frac{1}{k^2r^2}\left(
    \begin{array}{cccc}
      S_{11} & S_{12} & 0 & 0 \\
      S_{12} & S_{22} & 0 & 0 \\
      0 & 0 & S_{33} & S_{34} \\
      0 & 0 & -S_{34} & S_{44}
    \end{array}
  \right)
     \left(
    \begin{array}{cccc}
	I_{\rm inc} \\
	Q_{\rm inc}\\
	U_{\rm inc}\\
	V_{\rm inc}
    \end{array} 
    \right),
\end{equation}
where $k$ is the wave number, $r$ is the distance from the scatterer to the observer, and $S_{ij}$ ($i, j =$ 1, 2, 3, 4) represents the phase matrix elements. In general, a phase matrix contains 16 independent elements.
Assuming the rotational and mirror symmetry of the aggregates, the phase matrix has eight nonzero elements, six of which are independent \citep[][hereafter BH83]{bohren83}. 
Because of this symmetry, each phase matrix element depends on the scattering angle $\theta$ but not on the azimuthal angle $\phi$. Note that the scattering angle is defined as the angle between the incident wave vector $\vecbf{k_i}$ and the scattered wave vector $\vecbf{k_s}$.

\subsection{Scattering by fractal dust aggregates}
This section introduces three models for calculating the optical properties of fractal dust aggregates.
The term {\it monomer} is used to indicate a constituent particle of a dust aggregate, and for the sake of simplicity, all monomers are assumed to be spherical and identical throughout the aggregates.

\subsubsection{T-matrix method}
The optical properties of dust aggregates were calculated using the TMM, which is one of the most rigorous methods of calculating the optical properties of an ensemble of spheres \citep[for a review, see, e.g.,][]{mishchenko96}. 
By virtue of the exact Mie solution, the optical properties of a collection of spherical particles can theoretically be obtained using the superposition principle.
We used the Fortran 77 code SCSMFO1B.FOR developed by \citet{mackowski96}, which is designed to calculate the optical properties of multiple spheres.
In addition, to reduce the numerical expense, orientation averaging was conducted using the quasi-Monte Carlo method developed by \citet{okada08}. 
This averaging was performed with 30 orientations for each dust aggregate. A detailed model of the dust aggregates used in this study is described in Section \ref{sec:our}.

\subsubsection{Mie theory with effective medium approximation}
The EMT allows us to calculate the optical properties of dust aggregates using Mie theory \citep[e.g.,][]{chylek00}. 
Because Mie theory is only applicable to homogeneous spheres, each aggregate is replaced with a single homogeneous pseudosphere with an effective dielectric function. 
One way to obtain the effective dielectric function is to calculate the average polarizability of the vacuums and monomers weighted for the volume filling factor. 
The dielectric functions of the monomer (inclusion component) and vacuum (matrix component) are denoted by $\epsilon_{\rm i}$ and $\epsilon_{\rm m}$, respectively. The single inclusion embedded in the matrix has a polarizability of $\alpha_{0}=4\pi R_0^3 (\epsilon_{\rm i}-\epsilon_{\rm m})/(\epsilon_{\rm i}+2\epsilon_{\rm m})$, where $R_0$ is the radius of a monomer (see Equation (5.15) in BH83). 
The total polarizability of an aggregate of $N$ monomers can be presumed to be $\alpha_{\rm agg}=N\alpha_0$ as long as interactions between monomers can be disregarded.
Equating $\alpha_{\rm agg}$ with the polarizability of a sphere of radius $R_c$ and effective dielectric function $\epsilon_{\rm eff}$ yields
\begin{equation}
R_c^3\frac{\epsilon_{\rm eff}-\epsilon_{\rm m}}{\epsilon_{\rm eff}+2\epsilon_{\rm m}}=NR_0^3\frac{\epsilon_{\rm i}-\epsilon_{\rm m}}{\epsilon_{\rm i}+2\epsilon_{\rm m}}. \label{eq:MG}
\end{equation}
Solving Equation (\ref{eq:MG}) for $\epsilon_{\rm eff}$ yields
\begin{equation}
\epsilon_{\rm eff}=\epsilon_{\rm m}\frac{\epsilon_{\rm i}+2\epsilon_{\rm m}+2f(\epsilon_{\rm i}-\epsilon_{\rm m})}
{\epsilon_{\rm i}+2\epsilon_{\rm m}-f(\epsilon_{\rm i}-\epsilon_{\rm m})}, \label{eq:MGeff} 
\end{equation}
where $f$ is the volume fraction of inclusions in the matrix and is given by 
\begin{equation}
f=N\left(\frac{R_0}{R_c}\right)^3 \label{eq:filling}.
\end{equation}
This effective dielectric function is known as the Maxwell--Garnett law \citep{Maxwell-Garnett1904, bohren83}.
Thus, a fractal dust aggregate of $N$ monomers can be replaced with a single pseudosphere of radius $R_c$ and effective dielectric function $\epsilon_{\rm eff}$.
Consequently, the optical properties of this pseudosphere can be readily obtained using Mie theory. 
Because of the symmetry of the phase matrix arising from the spherical symmetry of a particle, the EMT always yields $S_{11}=S_{22}$ and $S_{33}=S_{44}$. 

Equation (\ref{eq:MGeff}) is only valid when $f\lesssim10\%$ and $X_0\lesssim 0.5$ \citep{kolokolova01}. Hence, when at least one of these conditions is not fulfilled, it is necessary to use another mixing rule, such as the Bruggeman rule \citep{Bruggeman35} when $f\gtrsim10\%$ or the extended EMT theory \citep{Stroud1978, Wachniewski86} when $X_0\gtrsim0.5$.
It is worth noting that fluffy dust aggregates have small $f$, meaning the EMT yields results similar to the Rayleigh--Gans solutions for a sphere (see Chap. 6 of BH83). 

\subsubsection{Rayleigh--Gans--Debye theory} \label{sec:RGD}
The basic idea of the RGD theory is as follows.
Assuming that multiple scattering inside the aggregates can be ignored, the light scattered by all of the monomers is superposed, taking into account the phase differences between light rays. 

The RGD theory assumes that the field inside the particle is approximately the same as the external incident field. 
This assumption is valid when the following conditions are satisfied:
\begin{eqnarray}
|m-1|&\ll& 1, \label{eq:rgdcond1}\\
2X_0|m-1|&\ll&1, \label{eq:rgdcond2}\\
2X_c|m_{\rm eff}-1|&\ll&1, \label{eq:rgdcond3}
\end{eqnarray}
where $m$ is the complex refractive index of a monomer and $X_0$ and $X_c$ are the size parameters of the monomers and the aggregates of characteristics radius $R_c$, respectively. The size parameter is defined as
\begin{equation}
X=kR=\frac{2\pi{R}}{\lambda},
\end{equation}
where $\lambda$ is the wavelength measured in a vacuum. 

The refractive indices are related to the dielectric function as $m=\sqrt{\epsilon}$, and thus $m_{\rm eff}$ can be calculated using Equation (\ref{eq:MGeff}). 
Equation (\ref{eq:rgdcond1}) requires the absence of the reflection of light by a monomer.
Equations (\ref{eq:rgdcond2}) and (\ref{eq:rgdcond3}) require the changes in the amplitude and phase of incident light to be negligible within aggregates \citep{bohren83}. 
These conditions correspond to the fact that aggregates can be regarded as ``almost transparent".

When above conditions are satisfied and multiple scattering can be considered negligible, the phase matrix elements of the aggregates reduce to
\begin{equation}
S_{ij, {\rm agg}}(\theta)=N^2S_{ij, {\rm mono}}(\theta)\mathcal{S}(\vecbf{q}), \label{eq:RGD}
\end{equation}
where $S_{ij, {\rm agg}}$ and $S_{ij, {\rm mono}}$ represent $S_{ij}$ for the aggregate and monomer, respectively \citep{botet97, sorensen01}, and $\mathcal{S}(\vecbf{q})$ is the structure factor. %defined in Equation (\ref{eq:wk}). 
A relation similar to Equation (\ref{eq:RGD}) can be obtained by analogy to the theory of scalar wave scattering (see Appendix \ref{sec:appA}).
Because multiple scattering is ignored, the phase difference between scattered light rays can be determined from the relative position vector of every pair of monomers in the aggregate. Hence, it is helpful to introduce the following two-point correlation function:
\begin{eqnarray}
g(\vecbf{u})&\equiv&\int n(\vecbf{r})n(\vecbf{r}-\vecbf{u})d\vecbf{r}, \label{eq:cordef}
\end{eqnarray}
where $\vecbf{r}$ is the position vector, $\vecbf{u}$ is the relative position vector between two locations inside the scatterer, and $n(\vecbf{r})$ is the normalized distribution function of monomers. The normalized distribution function $n(\vecbf{r})$ is defined as
\begin{eqnarray}
n(\vecbf{r})&=&N^{-1}\sum_{i=1}^{N}\delta(\vecbf{r}-\vecbf{r_i}),\label{eq:nnorm}\\
\int n(\vecbf{r})d\vecbf{r}&=&1,
\end{eqnarray}
where $\delta(\vecbf{r})$ is the Dirac delta function, $\vecbf{r_i}$ is the position vector of the $i$th monomer, and $N$ is the number of monomers.
The structure factor in Equation (\ref{eq:RGD}) can be expressed as the Fourier transform of the following two-point correlation function (Wiener--Khinchin theorem):
\begin{equation}
\mathcal{S}(\vecbf{q})\equiv\int g(\vecbf{u})e^{i{\scriptsize \vecbf{q}\cdot\vecbf{u}}}d\vecbf{u}, \label{eq:wk}
\end{equation}
where $\vecbf{q}=\vecbf{k_s}-\vecbf{k_i}$ is the scattering vector. 
As a result, $\mathcal{S}(\vecbf{q})$ can be understood as the power spectrum that characterizes the configuration of fractal dust aggregates.
Because $n(\vecbf{r})$ is normalized to unity, $\mathcal{S}(\vecbf{q})$ approaches unity when $\vecbf{q}\cdot\vecbf{u}\ll1$.

Equation (\ref{eq:RGD}) shows that once the statistical properties $N$ and $\mathcal{S}(\vecbf{q})$ of the aggregate and those $S_{ij, {\rm mono}}$ of the monomer are specified, $S_{ij, {\rm agg}}$ can be obtained. Because the monomer is assumed to be spherical, $S_{ij, {\rm mono}}$ is calculated using Mie theory. 

In the RGD theory (Equation (\ref{eq:RGD})), every phase matrix element is proportional to the structure factor, and thus the ratio of $S_{ij}/S_{11}$ is not dependent on the structure factor. 
This suggests that the degree of polarization $P=-S_{11}/S_{12}$ of the aggregates can be determined from that of their constituent particles (monomers).
For example, if the monomers show Rayleigh-like scattering, the degree of polarization of the aggregates shows similar behavior.
It is worth noting that these characteristics have been observed in both experiments \citep[e.g.,][]{Volten2007} and numerical simulations \citep[e.g.,][]{min15}.

\section{Dust model} \label{sec:dust}
As shown in Section \ref{sec:RGD}, once the statistical quantities of fractal aggregates, such as the two-point correlation function and the structure factor, have been obtained, their light scattering intensity at each angle can be determined. 
This section first introduces the radius of the aggregate. Second, an analytical expression describing the structure factor for the fractal aggregate is described. Finally, the aggregate models used in the calculation are summarized and compared with the analytic formula of the structure factor.

\subsection{Statistical properties of fractal dust aggregates}
Fractal dust aggregates do not have a specific configuration, but their configurations can be characterized statistically.
This section describes how fractal dust aggregates can be statistically characterized.

\subsubsection{Radius of fractal dust aggregates}
Two frequently used definitions of the radius of a dust aggregate is the radius of gyration $R_g$ and the characteristic radius $R_c$.
$R_g$ represents the dispersion of the mass of the monomers with respect to their center of mass, as \citep[e.g.,][]{Mukai92}
\begin{equation}
R_g=\left[\frac{1}{2N^2}\sum_i\sum_j(\vecbf{r}_i-\vecbf{r}_j)^2\right]^{1/2} \label{eq:rg}.
\end{equation}
Again, all monomers are assumed to be identical.
When the radius of gyration is calculated for homogeneous spheres of radius $a$, it reduces to $R_g=\sqrt{3/5}a$.
When the fractal aggregates are replaced with the pseudospheres used in the EMT calculations, it is convenient to define the characteristic radius, which is given by \citep[e.g.,][]{Mukai92}
\begin{equation}
R_c=\sqrt{\frac{5}{3}}R_g. \label{eq:rc}
\end{equation}

\subsubsection{Analytical expression of structure factor} \label{sec:analytic}
A model of the two-point correlation function for fractal dust aggregates is introduced as \citep[e.g.,][]{Teixeira86}
\begin{equation}
 g(\vecbf{u})=Au^{d_f-3}\exp\left[{-(u/\xi)^\beta}\right]+\frac{1}{N}\delta(\vecbf{u}), \label{eq:gumodel}
\end{equation}
where $A$ is a constant, $u=|\vecbf{u}|$, $\xi$ is the cut-off radius or correlation length, and $\beta$ represents the power of the cut-off.
In Equation (\ref{eq:gumodel}), the power law function characterizes the fractal structure, where $d_f$ represents the fractal dimension and the other exponent of 3 comes from the dimension of the space. 
The meaning of the fractal dimension can be understood as follows.
The number of monomers within a distance $R$ from a monomer is $N\propto\int_0^R g(u)4\pi{u^2}du\propto\int_0^R u^{d_f-3}4\pi{u^2}du\propto R^{d_f}$, where $R\ll\xi$ is assumed and the second term in Equation (\ref{eq:gumodel}) is ignored. 
This yields another expression for the fractal dimension:
\begin{equation}
N=k_0\left(\frac{R_g}{R_0}\right)^{d_f}, \label{eq:fractal}
\end{equation}
where the prefactor $k_0$ is a constant. Note that this expression holds when $N$ is sufficiently large.
The fractal dimension describes the dimension of the monomer distribution. For example, when the monomers are distributed with a fractal dimension of two, they are distributed as if they were in a two-dimensional space. 
BCCAs, which are the product of a series of mutual collisions between aggregates of comparable masses, tend to have $k_0\simeq1.04$ and $d_f\simeq1.9$ for offset collisions and $d_f\simeq2.0$ and $k_0\simeq1.03$ for head-on collisions. 
BPCAs, which are formed by sticking of monomers one by one, tend to have $k_0\simeq0.30$ and $d_f\simeq3.0$. 
The cut-off function in Equation (\ref{eq:gumodel}) is intended to take the aggregate size into account.
$\beta$ represents the power of the cut-off, but its value for fractal aggregates is controversial.
Owing to its mathematical simplicity, the exponential cut-off model ($\beta=1$) has been investigated by numerous authors \citep[e.g.,][]{sinha84, chen86, freltoft86, berry86,kozasa93, filippov00}. However, \citet{sorensen92} pointed out that light scattering experiments on fractal dust aggregates support the Gaussian cut-off model ($\beta=2$).
Section \ref{sec:our} will demonstrate that the Gaussian cut-off model more accurately reproduces actual BCCAs and BPCAs than the exponential model; thus, the Gaussian cut-off model ($\beta=2$) is adopted in the following discussion.
The second term in Equation (\ref{eq:gumodel}) is introduced to ensure this equation is consistent with Equation (\ref{eq:cordef}).
This term only becomes important when $N$ is small. The calculation of the TMM is confined to small $N$ because of the computational demand; hence, this term cannot be negligible.

The constant $A$ in Equation (\ref{eq:gumodel}) is determined by the unitary condition
\begin{equation}
\int g(\vecbf{u})d\vecbf{u}=1, \label{eq:norm1} 
\end{equation}
and the correlation length $\xi$ is determined from the following relationship \citep{sorensen01}:
\begin{equation}
R_g^2=\frac{1}{2}\int |\vecbf{u}|^2g(\vecbf{u})d\vecbf{u}. \label{eq:norm2}
\end{equation}
Equations (\ref{eq:gumodel}), (\ref{eq:norm1}), and (\ref{eq:norm2}) yield
\begin{eqnarray}
A&=&\frac{\beta}{4\pi\xi^{d_f}\Gamma(d_f/\beta)}\left(1-\frac{1}{N}\right), \label{eq:SorA}\\
\xi^2&=&\frac{2\beta}{(d_f-\beta+2)}\frac{\Gamma(d_f/\beta)}{\Gamma((d_f+2)/\beta-1)}\left(1-\frac{1}{N}\right)^{-1}R_g^2, \label{eq:Sorxi} \nonumber\\
\end{eqnarray}
where $\Gamma(z)$ is the Gamma function.

Assuming that the monomer distribution is isotropic and $\beta=2$,
it follows from Equations (\ref{eq:wk}) and (\ref{eq:gumodel}) that 
\begin{eqnarray}
\mathcal{S}(q)&=&4\pi A \int_0^{\infty} u^{d_f-1} e^{-(u/\xi)^2} \frac{\sin{qu}}{qu} du  + \frac{1}{N}\int \delta(\vecbf{u})e^{i{\scriptsize \vecbf{q}\cdot\vecbf{u}}}d\vecbf{u} \nonumber\\
&=&4\pi{A}\sqrt{\frac{\pi}{2q}} \int_0^{\infty} u^{d_f-\frac{3}{2}}e^{-(u/\xi)^2}J_{\frac{1}{2}}(qu) du + \frac{1}{N}, \nonumber \\
\end{eqnarray}
where $J_{\frac{1}{2}}(x)$ is a spherical Bessel function of the first kind. Using the integral formula \footnote{see, e.g., Equation (6.631.1) of \citet{Gradshteyn:2007aa}}:
\begin{eqnarray}
&&\int_0^{\infty} x^{\mu}e^{-\alpha{x^2}}J_{\nu}(\gamma{x})dx \nonumber\\
&=&\frac{\gamma^{\nu}\Gamma(\frac{\mu+\nu+1}{2})}{2^{\nu+1}\alpha^{\frac{\mu+\nu+1}{2}}\Gamma(\nu+1)}
{}_1F_1\left(\frac{\mu+\nu+1}{2}; \nu+1; -\frac{\gamma^2}{4\alpha}\right) \nonumber \\
\end{eqnarray}
finally yields
\begin{equation}
\mathcal{S}(q)=\left(1-\frac{1}{N}\right){}_1F_1\left(\frac{d_f}{2};\frac{3}{2};-\frac{(q\xi)^2}{4}\right)+\frac{1}{N}, \label{eq:gauss0}
\end{equation}
where ${}_1F_1$ is the confluent hypergeometric function (see Appendix \ref{sec:appB}) and $q=|\vecbf{q}|=2k\sin(\theta/2)$.
Once $R_g$, $N$, and $d_f$ are obtained, the structure factor of the fractal dust aggregates can be determined. It should be noted that Equation (\ref{eq:gauss0}) in this paper reduces to Equation (9) of \citet{sorensen92} in the limit of $N \to \infty$.
In the following discussion, for the sake of simplicity, $\xi^2\simeq4R_g^2/d_f$ is assumed, which yields
\begin{equation}
\mathcal{S}(q)\simeq\left(1-\frac{1}{N}\right){}_1F_1\left(\frac{d_f}{2};\frac{3}{2};-\frac{(qR_g)^2}{d_f}\right)+\frac{1}{N}. \label{eq:gauss}
\end{equation}
Note that the analytic form of $\mathcal{S}(q)$ for $\beta=1$ has been given in previous studies \citep[e.g.,][]{berry86}.

\subsubsection{Simplistic form of structure factor for large $qR_g$}
For $qR_g\gg1$, using the asymptotic form of the hypergeometric function yields
\begin{eqnarray}
{}_1F_1\left(\frac{d_f}{2};\frac{3}{2};-\frac{(qR_g)^2}{d_f}\right)\simeq\left\{ \begin{array}{ll}
C(qR_g)^{-d_f} & (d_f < 3) \\
\exp\left[-(qR_g)^2/d_f\right]  & (d_f=3), \\
\end{array} \right. \label{eq:simple}
\end{eqnarray}
where $C$ is a function of the fractal dimension and $C=1$ when $d_f=2$ (see Appendix \ref{sec:appB}).
In the case of BCCAs, using Equation (\ref{eq:fractal}) with $d_f=2$ and $k_0=1$ for simplicity yields
\begin{eqnarray}
\mathcal{S}(q)&\simeq& \left(1-\frac{1}{N}\right)(qR_0)^{-2}\frac{1}{N}+\frac{1}{N}\\
&\simeq&\frac{1}{N}\{(qR_0)^{-2}+1\},
\end{eqnarray}
where higher orders of $(1/N)^2$ are ignored. As a result, 
\begin{equation}
S_{ij, {\rm agg}}(\theta)\simeq{N}S_{ij, {\rm mono}}(\theta)[(qR_0)^{-2}+1]. \label{eq:enhance}
\end{equation}
The expression $[(qR_0)^{-2}+1]$ inside the square brackets in Equation (\ref{eq:enhance}) indicates the enhancement of the scattered intensity due to interference of the scattered waves of each monomer. This term may be important to distinguish between light scattered by separately distributed monomers and that scattered by fluffy dust aggregates with $d_f=2$. 
Because such aggregates are thought to be present at the surface layer of protoplanetary disks owing to their strong dynamical coupling to disk gas, Equation (\ref{eq:enhance})
 might be useful to model the scattered light of the disks. Note that this expression holds for large angle scattering, defined as $\theta_{\rm min}<\theta<\pi$, where $\theta_{\rm min}\propto X_{\rm agg}^{-1}$ (see Section \ref{sec:scalingap} for more detail). Thus, as the size of the aggregate increases, this expression becomes accurate for most scattering angles.

\subsection{Proposed particle models} \label{sec:our}
\begin{figure}[t]
\begin{center}
\includegraphics[height=5.0cm,keepaspectratio]{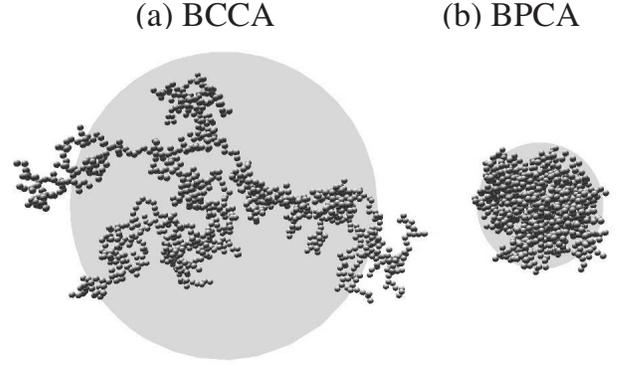}
\caption{(a) BCCA and (b) BPCA models. The number of monomers in each model is 1024, and the BCCA and BPCA models have radii $R_c/R_0$ of approximately 46.2 and 19.6, respectively. The gray shadowed regions illustrate the characteristic radii of the aggregates. }
\label{fig:config}
\end{center}
\end{figure}

This paper considers two types of dust aggregates, BCCAs and BPCAs. BCCAs are generated by a series of mutual collisions of aggregates of comparable masses, whereas BPCAs are generated by sticking of monomers one by one. %monomers sticking to each other one by one. 
BCCAs have a highly porous structure, whereas BPCAs have a more compact structure. 
BCCAs and BPCAs typically have fractal dimensions of $d_f \lesssim 2.0$ and $d_f \simeq 3.0$, respectively.
Simulations were performed to numerically model these aggregates. 
The number of monomers was set to $N=128, 256, 512$, $1024$, and $8192$. 
Figure \ref{fig:config} shows examples of the two types of aggregate models generated with $1024$ monomers.

\begin{figure}[b]
\begin{center}
\includegraphics[height=7.0cm,keepaspectratio]{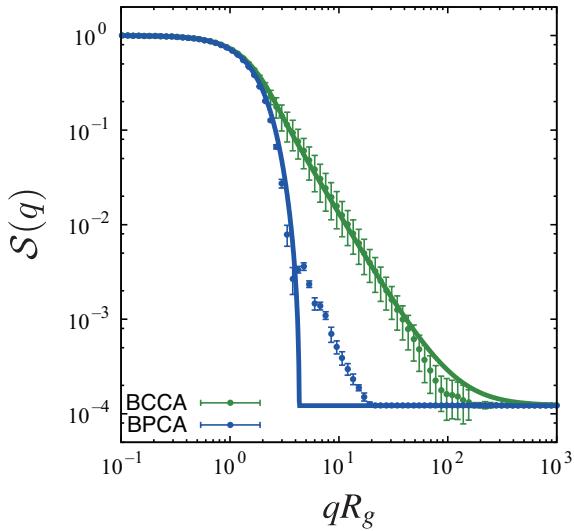}
\caption{Statistical models and actual values of structure factors for BCCAs (green) and BPCAs (blue) with $N=8192$. The radius of gyration and the fractal dimension are given in Table \ref{tab:stat}. Dots with error bars represent the actual structure factors of the aggregates, and solid lines indicate the structure factors obtained using the Gaussian correlation model calculated from Equation (\ref{eq:gauss}).}
\label{fig:statics}

\end{center}
\end{figure}

\begin{table*}
\caption{Statistical Quantities of Proposed BPCA and BCCA Models for Various Numbers of Monomers. }
\label{tab:r}
\centering
\begin{tabular}{llcllcllcllcll}
\hline
& ${R}_{\rm {V}}$\footnote{The radius of volume-equivalent spheres and the monomer radius are assumed to be $R_0=0.1\ \upmu$m.} &&\multicolumn{2}{c}{$R_{g,\ {\rm ours}}$
[$\upmu$m]} &&\multicolumn{2}{c}{$R_{g,\ {\rm asymptotic}}$
[$\upmu$m]} && \multicolumn{2}{c}{$d_f$ ($\beta=1$)} && \multicolumn{2}{c}{$d_f$ ($\beta=2$)} 
 \\
  \cline{4-5}\cline{7-8}\cline{10-11}\cline{13-14}
$N$  & [$\upmu$m] && BPCA & BCCA && BPCA & BCCA && BPCA\footnote{A best fit value could not be found in the range $1.5<d_f<4.0$.} & BCCA && BPCA & BCCA \\
\hline  \hline
$128$ &$0.50$&&$0.73$&$1.28$&&$0.75$&$1.26$ &&$-$ & $3.13$&&$3.12$ & $1.99$\\
$256$ &$0.63$&&$0.95$&$1.92$&&$0.95$&$1.81$ &&$-$ & $2.77$&&$2.95$ & $1.85$\\
$512$ &$0.80$&&$1.23$&$2.62$&&$1.20$&$2.61$ &&$-$ & $2.72$&&$2.95$ & $1.88$\\
$1024$ &$1.0$&&$1.51$&$3.73$&&$1.51$&$3.76$ &&$-$ & $2.60$&&$3.00$ & $1.91$\\
$8192$ &$2.02$&&$3.05$&$11.4$&&$3.01$&$11.2$ &&$-$ & $2.30$&&$3.06$ & $1.89$\\
\hline
\end{tabular}
\label{tab:stat}
\end{table*}

Even for models with the same number of monomers and agglomerate type, the aggregate configuration can vary widely because of the randomness of the collisional parameters, such as the impact parameter and the orientation of the aggregates. To remove the effect of the randomness, $N_a$ aggregates were produced for $N=128, 256, 512$, $1024$, and $8192$, where $N_a$ is the number of statistically independent aggregates. For all values of $N$ except $N=8192$, $N_a=10$ and $4$ were adopted for the BCCA and BPCA models, respectively. For $N=8192$, $N_a=100$ and $10$ were adopted for the BCCA and BPCA models, respectively.

First, the radii of gyration of our generated aggregates are described.
The radii of gyration of the generated aggregates were calculated using Equation (\ref{eq:rg}), and the arithmetic mean value among $N_a$ aggregates is given in Table \ref{tab:stat}.
For comparison, the radii of gyration were also calculated using Equation (\ref{eq:fractal}), which gives their asymptotic values for large $N_a$. In Equation (\ref{eq:fractal}), the fractal dimensions of BCCAs and BPCAs are assumed to be $d_f=1.9$ and $3.0$, respectively, and the prefactors adopted for the BCCA and BPCA models are $k_0=1.04$ and $0.30$, respectively. The generated aggregates have radii of gyration that are almost equal to their asymptotic values for large $N_a$. It should be noted that the small values of $N_a$ are likely responsible for the slight discrepancy between the radii of the aggregates and their asymptotic values. The following discussion adopts the radius of gyration calculated using Equation (\ref{eq:rg}).

Second, the fractal dimensions of the aggregates were determined by employing two different models of correlation functions.
The fractal dimension can be determined from the chi-square fitting of the measured correlation function using Equation (\ref{eq:gumodel}).
The results of the fits are summarized in Table \ref{tab:stat}. Table \ref{tab:stat} shows that the Gaussian cut-off model ($\beta=2$) is consistent with the typical fractal dimensions of BCCAs and BPCAs, whereas the exponential cut-off model ($\beta=1$) is not. 
Indeed, the reduced chi-square values for $\beta=1$ and $2$ for the BCCA model containing 8192 monomers are $\chi_\nu^2=3.98$ and $2.20$, respectively, and therefore the Gaussian cut-off model ($\beta=2$) exhibits a better fit. 
A suitable $d_f$ could not be found within the range $1.5<d_f<4.0$ for BPCAs when the exponential cut-off model ($\beta=1$) was employed.
As a result, the Gaussian cut-off model was adopted in Equation (\ref{eq:gumodel}). It should be noted that the obtained reduced chi-square for $\beta=2$ is slightly larger than 1.0 because of a strong peak at $u=2R_0$ and a discontinuity at $u=4R_0$ \citep[see, e.g., Figure 2 of ][]{hasmy93}.

Figure \ref{fig:statics} shows a plot of the structure factor calculated using the proposed particle model and statistical model.
Small $qR_g$ values indicate large-scale structures, whereas large $qR_g$ values represent small-scale structures. 
Figure \ref{fig:statics} indicates that the modeled structure factor reproduces the measured statistical properties of fractal dust aggregates in most $qR_g$ regimes. Although Equation (\ref{eq:gauss}) yields nearly correct results for the BCCA model, it fails for the BPCA model at small scales.

The slope of the actual structure factor for BPCAs is similar to the slope for BCCAs at $4\lesssim qR_g \lesssim 20$.
The main reason is that BCCAs and BPCAs are hardly distinguishable at small scales for $u\lesssim8R_0$.

\section{Results} \label{sec:results}
\begin{figure}[t]
\begin{center}
\includegraphics[height=7.0cm,keepaspectratio]{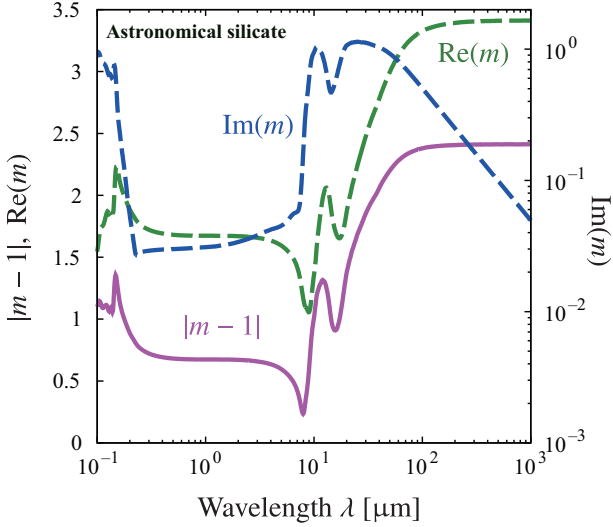}
\caption
{Optical constants of astronomical silicate. }
\label{fig:m-1}
\end{center}

\end{figure}

In this study, the chemical composition of the monomers was assumed to be the same as that of astronomical silicate \citep{draine84, laor93}, and the monomer radius was set to $R_0=0.1\upmu$m. 
Figure \ref{fig:m-1} shows a plot of the optical constants of astronomical silicate.
The phase matrix elements of fractal dust aggregates were calculated using the TMM, the RGD theory, and the EMT.

\subsection{Scattered intensity $S_{11}$} \label{sec:S11}
This section discusses the scattered intensity $S_{11}$ of unpolarized incident light scattered by fractal dust aggregates.

\subsubsection{Dependence on wavelength} \label{sec:wldep}
\begin{figure}[t]
\begin{center}
\includegraphics[height=7.0cm,keepaspectratio]{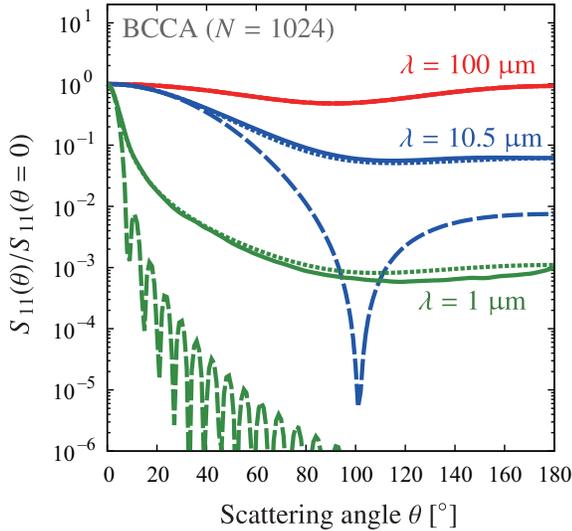}
\caption{$S_{11}$ of BCCA model with $N=1024$ for various wavelengths. 
Solid, dotted, and dashed lines show the results obtained using the TMM, the RGD theory, and the EMT, respectively. 
Red, green, and blue lines indicate the normalized intensities for incident wavelengths of $\lambda=100$, 10.5, and 1 $\upmu$m, respectively. In the case of $\lambda=100\ \upmu$m, the three different lines representing the three methods overlap each other. }
\label{fig:wldep}
\end{center}
\end{figure}
The wavelength dependence of the scattering is first discussed.
In Figure \ref{fig:wldep}, the scattered intensity normalized by $S_{11}(\theta=0)$ is plotted against the scattering angle $\theta$. The dust model is the BCCA model with $N=1024$, which means the radius of gyration is $R_g\simeq 3.73\ \upmu$m (see Table \ref{tab:r}).
If the size parameter $X_{\rm agg}=2\pi{R_g}/\lambda$ is much smaller than unity, the scattering can be understood in terms of Rayleigh or isotropic scattering. If the size parameter is larger than unity, the forward scattering intensity dominates the backward scattering even though each monomer scatters isotropically. As will be discussed in Section \ref{sec:angdep}, the underlying physics of the wavelength dependence can be clearly understood in terms of the interference between the light rays scattered by different monomers. 

Figure \ref{fig:wldep} also shows that the RGD theory is consistent with the wavelength dependence of the scattering results obtained using the TMM. Conversely, the EMT yielded accurate results for $X_{\rm agg}\ll1$ but not for $X_{\rm agg}>1$.

\subsubsection{Forward and backward scattering intensity}  \label{sec:angdep}
\begin{figure*}[htbp]
\begin{center}
\includegraphics[height=7.0cm,keepaspectratio]{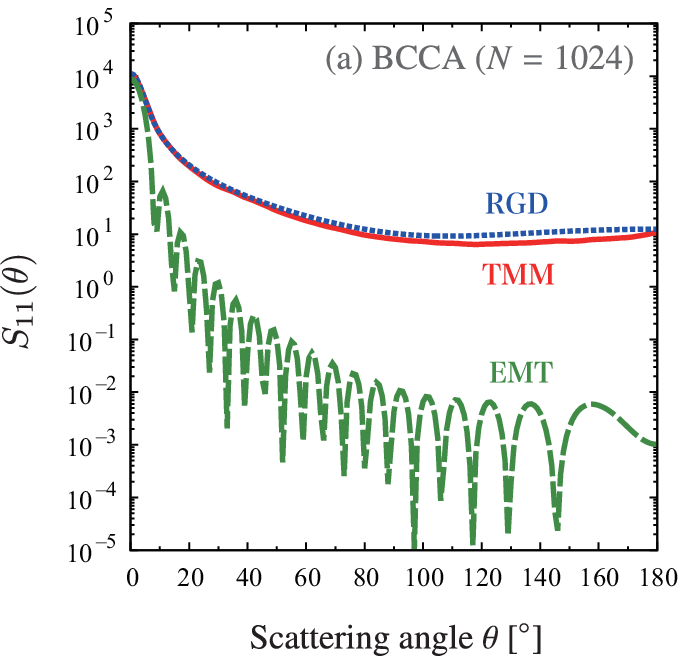}
\includegraphics[height=7.0cm,keepaspectratio]{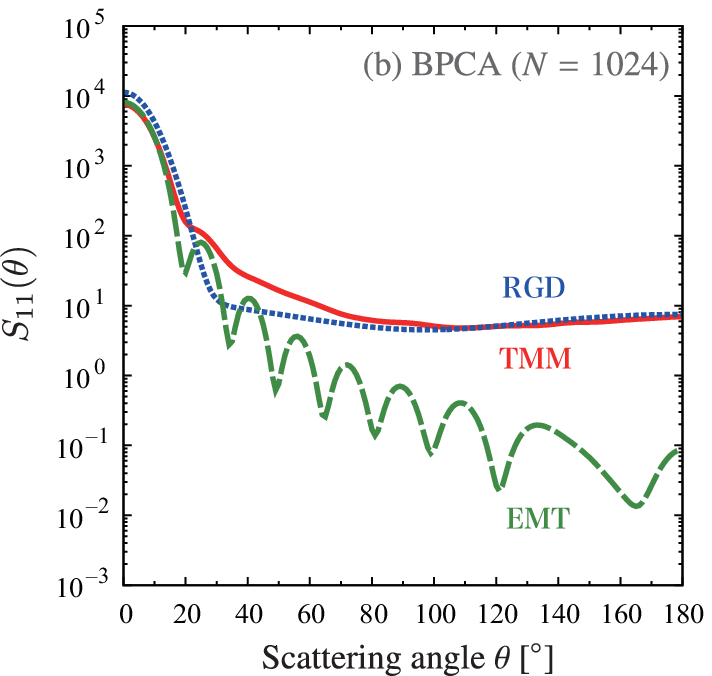}
\caption{$S_{11}$ of (a) BCCA and (b) BPCA models with $N=1024$. The red solid lines represent the rigorous results obtained using the TMM, and the blue dotted and green dashed lines represent the results obtained by the RGD theory and the EMT, respectively. The incident wavelength was set to $\lambda=1\ \upmu$m.}
\label{fig:1024}
\end{center}
\end{figure*}

\begin{figure}
\begin{center}
\includegraphics[height=7.0cm,keepaspectratio]{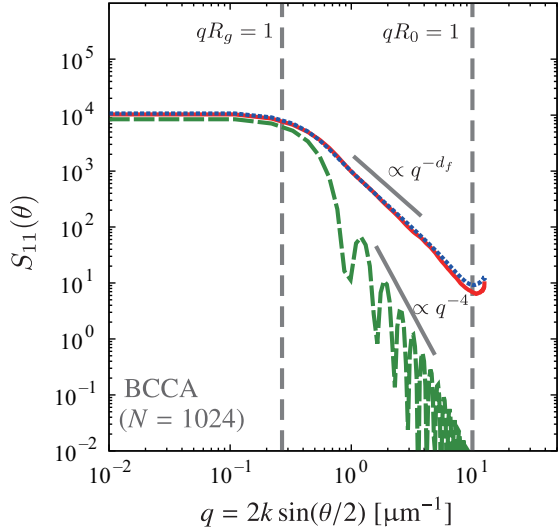}
\caption{$S_{11}$ of BCCA model containing $N=1024$ monomers plotted against $q=2k\sin(\theta/2)$. The red solid lines represent the rigorous results obtained using the TMM, and the blue dotted and green dashed lines represent the results obtained by the RGD theory and the EMT, respectively. The incident wavelength was set to $\lambda=1\ \upmu$m.  The thick gray solid line indicates the slope of the scattering for fractal aggregates with $d_f=1.91$ and solid spheres. The vertical gray dotted lines on the left and right indicate where $qR_g=1$ and $qR_0=1$, respectively. }
\label{fig:scalingtmm}
\end{center}
\end{figure}

This section investigates the reason for the intense forward scattering by an aggregate with $X_{\rm agg} >1$. 
Figure \ref{fig:1024}(a) shows the scattered intensity of the BCCA model with $N=1024$ for an incident radiation of $\lambda=1\ \upmu$m. This plot demonstrates that the RGD theory is in good agreement with the rigorous TMM results. For this reason, it is helpful to use Equation (\ref{eq:RGD}) to investigate the origin of the intense forward scattering.
By definition, forward scattering always gives rise to $q=0$, and then $\mathcal{S}=1$ (see Equations (\ref{eq:wk}) and (\ref{eq:norm1})).
Thus, Equation (\ref{eq:RGD}) gives $S_{11, \rm{agg}}(\theta=0^{\circ})=N^2S_{11, {\rm mono}}(\theta=0^{\circ})$.
The reason the forward scattered intensity is proportional to $N^2$ is because of the coherent scattering. This can be intuitively understood as follows. In the forward scattering region, where $\vecbf{q}\cdot\vecbf{u}\ll1$, the waves scattered by each monomer are in phase and are thus added constructively. This means that the forward scattering amplitude is proportional to $N$; thus, the intensity is proportional to $N^2$. 
In the backward scattering region, where $\vecbf{q}\cdot\vecbf{u}\gg1$, Equation (\ref{eq:RGD}) reduces to $S_{11, \rm{agg}}(\theta=180^{\circ})=NS_{11, {\rm mono}}(\theta=180^{\circ})$ because $\mathcal{S}\to 1/N$ for $X_{\rm agg}\gg1$. 
In this case, the backward scattered intensity is proportional to $N$ because of the incoherent scattering. For backward scattering angles, the phases of the waves scattered by each monomer are random, meaning the waves are added destructively. For this reason, the amplitude is proportional to $\sqrt{N}$, and thus the intensity is proportional to $N$.

Figure \ref{fig:1024}(b) shows a plot of the scattered intensity for the BPCA model with $N=1024$.
As with BCCAs, BPCAs show intense forward scattering; however, they exhibit a slightly smaller forward scattered intensity than predicted by the RGD theory. 
This is related to the breakdown of the assumption of the RGD theory, which is discussed in more detail in Section \ref{sec:app}. 
The RGD theory also fails to reproduce accurate results at angles in the range of $20^{\circ}\lesssim\theta\lesssim100^{\circ}$. This inaccuracy comes from the error in the structure factor (see Figure \ref{fig:statics}).

\subsubsection{Dependence on scattering angle} \label{sec:scalingap} 
\begin{figure*}
\begin{center}
\includegraphics[height=6.5cm,keepaspectratio]{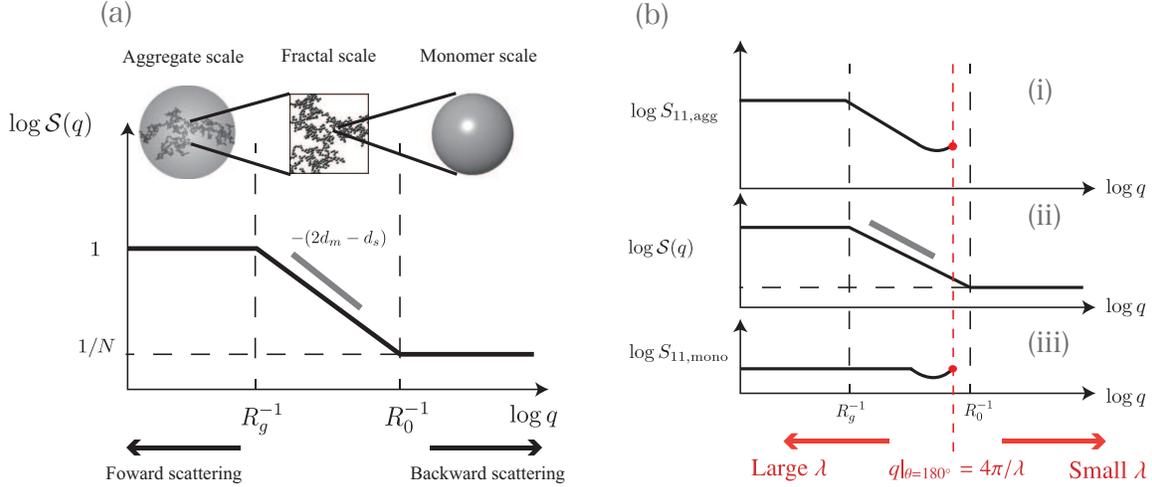}
\caption{(a) Schematic illustration of structure factor. The angular dependence of the scattering reflects the hierarchical structure of dust aggregates from large-scale structures (small $q$) to small-scale structures (large $q$). The slope of the fractal scale is determined by the surface fractal dimension $d_s$ and the mass fractal dimension $d_m$. (b) (i) Scattering by fractal dust aggregates, which can be determined by a combination of (ii) the structure factor and (iii) the scattering by the monomers. The monomers are assumed to be Rayleigh scatterers, meaning $R_0\lesssim\lambda/2\pi$. The smallest scale that appeared in the scattering pattern was $q=4\pi/\lambda$ at $\theta=\pi$.}
\label{fig:scaling}
\end{center}
\end{figure*}

Next, the angular distribution of the scattered intensity is discussed. Figure \ref{fig:scalingtmm} shows the scattered intensity plotted against the magnitude $q$ of the scattering vector. Figure \ref{fig:scalingtmm} shows that the angular distribution of the scattering can be scaled using $q$. This is because the structure factor or the power spectrum governs the scattering phenomenon. Equation (\ref{eq:RGD}) indicates that the angular distribution of the scattered light is determined from a combination of the phase matrix elements of a single monomer and the structure factor. The angular distribution of scattering is classified into the three regimes, the {\it aggregate}, {\it fractal}, and the {\it monomer scales}, according to the structure factor, as illustrated in Figure \ref{fig:scaling}(a).

The region in which $q<R_g^{-1}$ corresponds to the large-scale structure of aggregates and is termed the {\it aggregate scale}. 
In this regime, the detailed structure of the aggregate is not important, and the size of the aggregate plays a significant role.
Because the radius of a pseudosphere is intended to be equal to that of the aggregate in the EMT calculation, this calculation can produce accurate results in this regime. In other words, in this regime, the aggregate scatters light as if it were a single sphere with $R_g$. 
The value of $q$ decreases with decreasing scattering angle, and $q=0$ when $\theta=0$; hence, small-angle scattering is characterized by the large-scale structure of dust aggregates. Physically, this behavior is the result of the fact that small-angle scattering is coherent scattering.
It is worth noting that the aperture angle of the forward scattering, or primary, lobe can be characterized by $qR_g=1$, which yields $\theta_a\propto X_{\rm agg}^{-1}$, where $\theta_a$ is the aperture angle of intense forward scattering. As the size parameter increases, the forward scattering is focused into a narrower region. 

In the regime where $R_g^{-1}<q<R_0^{-1}$, the arrangement of monomers comes into play because the scattering is no longer coherent.
Because the arrangement of monomers can be described by the fractal dimension, the angular dependence of the scattered intensity can also be characterized by the fractal dimension. In this regard, this regime is denoted the {\it fractal scale}. 
In this regime, the scattered intensity is proportional to $q^{-d_f}$ (see Equation (\ref{eq:simple})), as demonstrated by Figure \ref{fig:scalingtmm}. In general, the slope is determined by the surface fractal dimension $d_s$ and the mass fractal dimension $d_m$ as $-(2d_m-d_s)$ \citep[e.g.,][]{sorensen01}. In the case of fractal aggregates, $d_f=d_m=d_s$, and thus the slope is simply $-d_f$. In the case of a solid sphere composed of almost transparent material (Rayleigh--Gans sphere), the slope in the fractal regime equals $-4.0$ because $d_m=3$ and $d_s=2$. This is known as Porod's law \citep{porod1951}. 
Because the pseudosphere of the EMT for the BCCA model is almost the same as the Rayleigh--Gans sphere, it obeys Porod's law, as shown in Figure \ref{fig:scalingtmm}. As a result, the EMT for the BCCA model cannot yield the angular dependence of fractal dust aggregates in this regime. 

Increasing $q$ such that $q>R_0^{-1}$, the scattering becomes dominated by the small-scale structure, that is, monomers. Because of this, this regime is named the {\it monomer scale}. In this regime, the structure factor again has a constant value, and the optical properties of the monomer determine the scattering properties. Because the EMT calculation assumes an infinitesimally small monomer radius or infinitely large monomer number, at a fixed $R_g$, it fails to reproduce the angular dependence in this regime.

The smallest scale appearing in the scattering pattern is roughly comparable to the wavelength of the incident light.
If $\lambda\gtrsim R_0$, the monomer scale is not relevant to the scattering pattern, and the backward scattering can be interpreted as Rayleigh scattering from the volume within a radius of $\lambda$ (see Figure \ref{fig:scaling}(b)). Therefore, the incident wavelength $\lambda$ functions as the spatial resolution of the structure.

The angular dependence of the scattering reveals the hierarchical structure of fractal aggregates from large-scale structures in the forward direction to small-scale structures in the backward direction. 
The reason the EMT calculation fails to reproduce the backscattering regime (fractal and monomer scales) is because small-scale structures, such as the arrangement of the monomers and the structure of the monomer itself, are ignored.
It is worth noting that a BPCA model containing a large number of monomers may not show a hierarchical scattering structure, because the hierarchy is based on the RGD theory, which is not applicable to large BPCA models because of the inaccuracy of the assumptions (see Section \ref{sec:app} for more detail). 

\subsubsection{Dependence on monomer number}
\begin{figure*}
\begin{center}
\includegraphics[height=7.0cm,keepaspectratio]{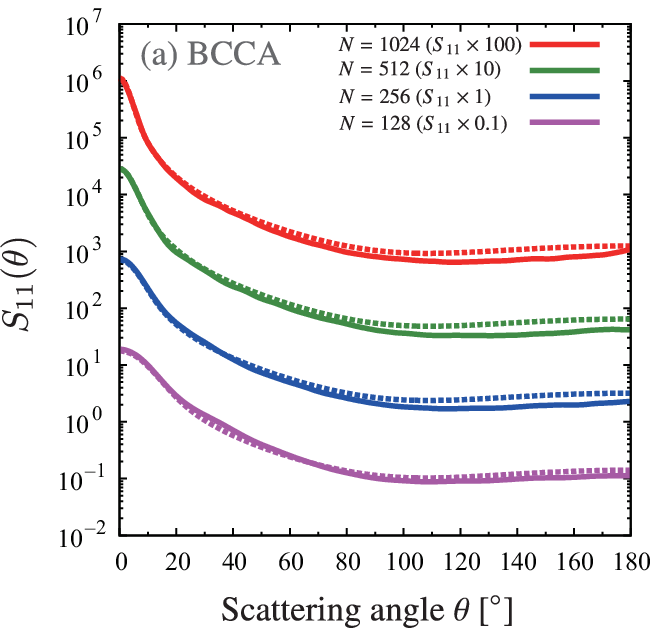}
\includegraphics[height=7.0cm,keepaspectratio]{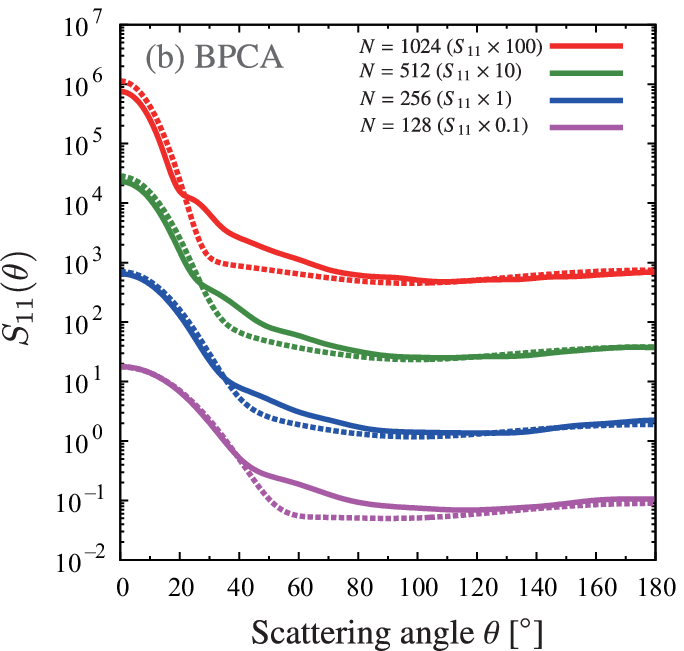}
\caption{$S_{11}$ of (a) BCCA and (b) BPCA models with $N=128, 256, 512$, and $1024$. Solid and dashed lines indicate the results obtained using the TMM and the RGD theory, respectively. To allow each curve to be easily distinguished, $S_{11}$ was artificially multiplied by $0.1, 1, 10$, and $100$ for $N=128, 256, 512$, and $1024$, respectively. The incident wavelength was set to $\lambda=1\ \upmu$m.}
\label{fig:Ndep}
\end{center}

\end{figure*}

Figure \ref{fig:Ndep} shows the dependence of the scattered intensity on the number of monomers in the BCCA and BPCA models.
Again, the incident wavelength was set to $1.0\ \upmu$m.
For both dust models, as the monomer number increases, the forward and backward scattering intensities also increase as a consequence of the interference. 
For the BCCA model, even for high numbers of monomers, the RGD theory still achieves high accuracy.
This is because the phase shift of the incident light within BCCAs remains nearly constant with increasing $N$, whereas it varies in BPCAs.
In addition, multiple scattering is not relevant to the aggregates if $d_f\leq2$ \citep{berry86, botet97}; therefore, the condition for the RGD is still satisfied as $N$ increases. However, for large $N$, the effect of multiple scattering cannot be ignored for $d_f>2$, as is the case in BPCAs. 

\subsection{Degree of linear polarization} \label{sec:LDP}
\begin{figure}[t]
\includegraphics[height=6.0cm,keepaspectratio]{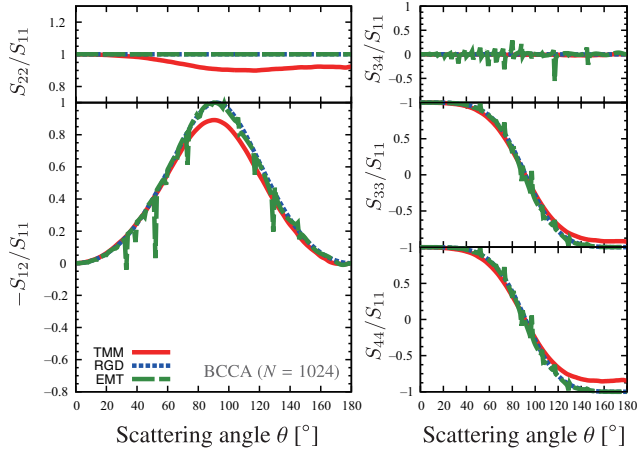}
\caption{Phase matrix elements $S_{12}, S_{22}, S_{33}, S_{34}$, and $S_{44}$ normalized by $S_{11}$ for the BCCA model with $N=1024$. The red line represents the rigorous results obtained using the TMM. The blue and green lines represent the results obtained using the RGD theory and the EMT, respectively. 
The incident wavelength was set to $\lambda=1\upmu$m. }
\label{fig:PMEcc}
\end{figure}

\begin{figure}[t]
\includegraphics[height=6.0cm,keepaspectratio]{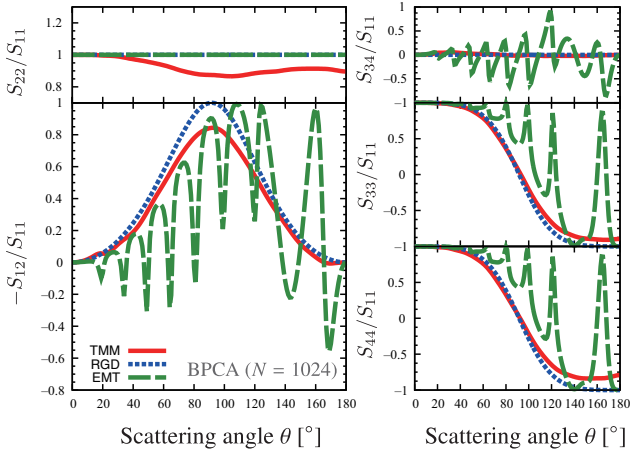}
\caption{Same as Figure \ref{fig:PMEcc} but for the BPCA model.}
\label{fig:PMEpc}
\end{figure}
This section discusses the degree of polarization of the aggregates.
The degree of linear polarization $P$ is defined as $P=-S_{12}/S_{11}$. 
The bottom left graphs in Figures \ref{fig:PMEcc} and \ref{fig:PMEpc} show plots of the degrees of linear polarization as functions of the scattering angle for the BCCA and BPCA models, respectively, with $N=1024$.
The rigorous calculation using the TMM shows that the maximum $P$ is achieved for $\theta\simeq90^\circ$ and that the angular distribution is almost symmetric about $\theta\simeq90^\circ$. 
As predicted by the RGD theory (see Section \ref{sec:RGD}), the degrees of linear polarization for both BCCAs and BPCAs exhibit angular distributions similar to that of a monomer, which is the Rayleigh scatterer.
However, the maximum $P$ obtained in the rigorous TMM results was slightly smaller than $100\%$.
In the case of a spherical grain, depolarization occurs when the size parameter exceeds unity. 
However, the mechanism of depolarization for the aggregate is essentially different from that for a spherical grain.
This aggregate depolarization is due to the occurrence of cross-polarization. Appendix \ref{sec:crosspol} discusses depolarization by cross-polarization in more detail.

Thus, the RGD theory can achieve a symmetric angular distribution of the degree of linear polarization of fractal aggregates with respect to $\theta=90^\circ$ but fails to reproduce its magnitude by a small margin. 
In the case of the EMT for the BCCA model, the solution is similar to the Rayleigh--Gans solution (see Chap. 6 in BH83), and the polarization is equal to the Rayleigh scattering.
Note that the EMT results show some spiky features at certain angles; these originate from the small but non-negligible phase shift of the incident light by a pseudosphere. When the phase shift is negligible, this spiky feature does not appear in $P$; however, this feature arises as the phase shift approaches unity. This spiky feature develops a more wavy pattern with increasing phase shift, as shown in the bottom left plot in Figure \ref{fig:PMEpc}.

\subsection{Other phase matrix elements}
The phase matrix elements $S_{12}, S_{22}, S_{33}, S_{34}$, and $S_{44}$ normalized by $S_{11}$ for the BCCA and BPCA models are shown in Figures \ref{fig:PMEcc} and \ref{fig:PMEpc}, respectively.
This section discusses elements $S_{33}, S_{34}$, and $S_{44}$ of the phase matrix, and Appendix \ref{sec:crosspol} discusses $S_{22}$. In the case of Rayleigh scattering, that is, in the Rayleigh--Gans solution, $S_{34}$ vanishes (see Chap. 13 of BH83), and thus $S_{34}$ in the RGD theory also vanishes.

Although some spikes can be observed in the polarization pattern because of the non-negligible amount of phase shift, $S_{34}$ in the EMT for the BCCA model is almost zero.
$S_{34}$ in the EMT for the BPCA model shows more complex behavior because the BPCA model is not transparent.
$S_{33}$ and $S_{44}$ are the same as those of a monomer in the absence of cross-polarization, and the angle dependence becomes $S_{33}/S_{11}=S_{44}/S_{11}=2\cos\theta/(1+\cos^2\theta)$ (see Equation (5.5) of BH83).
The results obtained using the RGD theory follow this formula. As shown in Figures \ref{fig:PMEcc} and \ref{fig:PMEpc}, at large scattering angles, the rigorous TMM results deviate from those obtained using the RGD theory. 
This inconsistency may be attributable to the occurrence of cross-polarization (Appendix \ref{sec:crosspol}).

\section{Applicability of RGD theory} \label{sec:app}
Section \ref{sec:results} demonstrated that the RGD theory can achieve results that are in fairly good agreement with the rigorous TMM results for most phase matrix elements and that this is true especially for BCCA models. 
As described in Section \ref{sec:RGD}, the conditions for the RGD theory are given by Equations (\ref{eq:rgdcond1}), (\ref{eq:rgdcond2}), and (\ref{eq:rgdcond3}). 
This section discusses the applicability of the RGD theory to fractal dust aggregates.

\subsection{Relative error in scattered intensity} \label{sec:relaerr}
To estimate the relative error between the RGD theory and TMM results, the relative error $\Delta$ was defined as
\begin{equation}
\Delta=\frac{S_{11}({\rm RGD})-S_{11}({\rm TMM})}{S_{11}({\rm TMM})},
\end{equation}
where $S_{11}$ is evaluated at $\theta=0$.

\subsubsection{Relative error due to refractive indices}
\begin{figure}[t]
\begin{center}
\includegraphics[height=7.0cm,keepaspectratio]{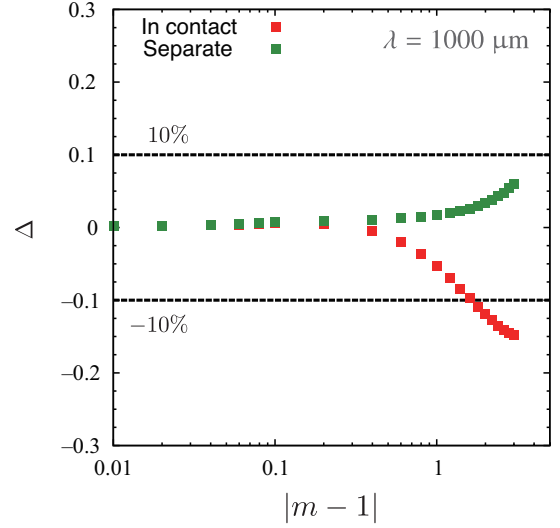}
\caption
{Relative error $\Delta$ between forward scattered intensities obtained using the RGD theory and the TMM plotted against refractive index.
The red dots indicate the dust aggregate with $R_g=3.61\ \upmu$m and $R_0=0.1\ \upmu$m, and a BCCA model with $N=1024$ is assumed; in this model, all monomers are in contact with each other.
The green dots represent the tentative dust model; in this model, each monomer is separated from its nearest neighbor(s) by a distance of $2R_0$.
The considered refractive indices range from $|m-1|=0.01$ to $3$. 
The incident wavelength was set to $\lambda=1000\ \upmu$m. The upper and lower horizontal lines represent errors of $10\%$ and $-10\%$, respectively.}
\label{fig:error2}
\end{center}

\end{figure}
Figure \ref{fig:m-1} shows that astronomical silicate violates or only marginally satisfies Equation (\ref{eq:rgdcond1}). % depending on wavelength.
To investigate the error arising from large $|m-1|$ values, the relative error is shown as a function of the refractive index $|m-1|$ in Figure \ref{fig:error2}.
The incident wavelength was set to $\lambda=1000\ \upmu$m; hence, the angular dependence of every scattering matrix elements is equal to that of Rayleigh scattering (see Figure \ref{fig:wldep}). Figure \ref{fig:error2} shows that the error increases with increasing $|m-1|$ and that the RGD theory slightly underestimates the scattered intensity.
For example, when $|m-1|\lesssim0.5$, the RGD theory results agree with the TMM results to an accuracy of $\lesssim1\%$, whereas when $|m-1|=3$, the relative error increases to $\Delta\approx-15\%$.
The enhancement observed in the TMM calculation at large $|m-1|$ has been reported when using the DDA \citep[see, e.g., Figures 3 and 4 of ][]{kozasa92}, noting that in their paper, enhancement was observed in the absorption and scattering cross sections.
A possible explanation for this behavior is the monomer--monomer interactions. 
To test this possibility, all monomers were tentatively separated from each other by an interval of $2R_0$ in the BCCA model with $N=1024$; hence, the monomers were not in contact with each other in this tentative dust model. The error for the tentative dust model is plotted as green dots in Figure \ref{fig:error2}. The enhancement did not appear in the tentative dust model, and the RGD theory results were shown to agree with the TMM results to an accuracy of $\lesssim10\%$. Therefore, the enhancement of the scattered intensity at large $|m-1|$ could be interpreted as being a result of monomer--monomer interactions.

\subsubsection{Relative error due to phase shift}
\begin{figure}[t]
\begin{center}
\includegraphics[height=7.0cm,keepaspectratio]{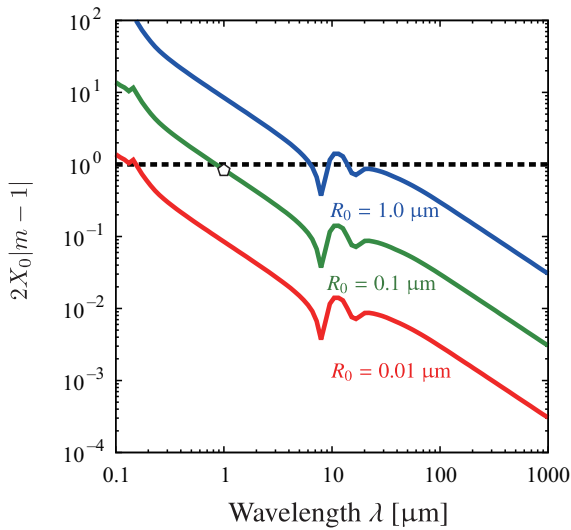}
\caption
{Phase shift induced by a single monomer. The red, green, and blue lines represent monomer radii of $R_0=0.01$, 0.1, and $1.0\ \upmu$m, respectively. The pentagon symbol represents a monomer with $R_0=0.1\ \upmu$m at $\lambda=1\ \upmu$m.}
\label{fig:phase0}
\end{center}
\end{figure}

The phase shift induced by a single monomer is plotted as a function of incident wavelength in Figure \ref{fig:phase0}.
The phase shift for a monomer with $R_0=0.1\ \upmu$m at $\lambda=1\ \upmu$m is $2X_0|m-1|\sim 0.8$. Thus, the condition given by Equation (\ref{eq:rgdcond2}) is marginally satisfied, 
although left-hand side approaches unity as the wavelength decreases. 
Figure \ref{fig:phase0} shows that the condition given by Equation (\ref{eq:rgdcond2}) is only marginally satisfied for large monomers at short wavelengths. Note that the phase shift depends on the composition of the monomers. For example, this condition is only marginally satisfied for opaque materials, such as graphite. 

\begin{figure}[t]
\begin{center}
\includegraphics[height=7.0cm,keepaspectratio]{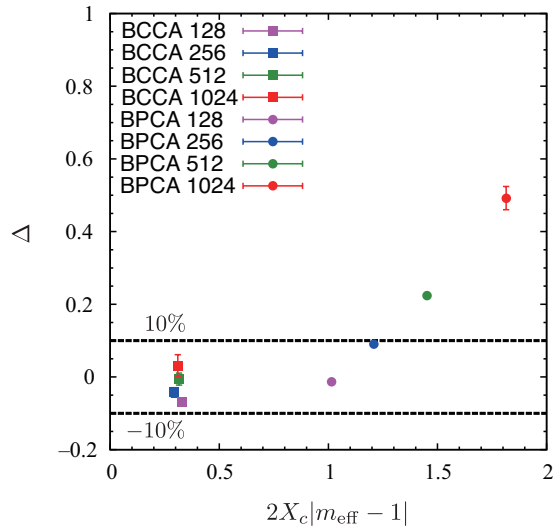}
\caption
{Relative error $\Delta$ between the RGD theory and the TMM results as a function of phase shift caused by aggregates. The squares and circles represent the BCCA and BPCA models, respectively. The magenta, blue, green, and red data points represent $N=128, 256, 512$, and $1024$, respectively. The incident wavelength was set to $\lambda=1\ \upmu$m. The upper and lower horizontal lines represent errors of $10\%$ and $-10\%$, respectively.} 
\label{fig:error}
\end{center}
\end{figure}

Figure \ref{fig:error} shows the relative error $\Delta$ plotted against the phase shift for different aggregate models and incident light with $\lambda=1\ \upmu$m.
The dependence of the phase shift of the BCCA model on $N$ is weak. Because $|m_{\rm eff}-1|\approx f|m-1|$, the phase shift is proportional to $R_cf$. Equation (\ref{eq:fractal}) can be used to demonstrate that the phase shift by the aggregates is proportional to $N^{1-{2/d_f}}$. Thus, in the case of $d_f=2$, as in BCCAs, the phase shift by aggregates does not depend on the number of monomers. 

Therefore, the relative error of the BCCA model is expected to be independent of $N$. However, as shown in Figure \ref{fig:error}, the relative error in the BCCA model grows gradually with increasing $N$. This increasing error may be attributable to the overlapping of monomers along the line of slight. This argument assumes that the prefactor $k_0$ does not depend on $N$, but this is not generally true. \citet{minato06} studied the projected area in BCCAs and BCCAs and found an empirical formula for this.
If an empirical formula can be extrapolated to large $N$, the degree of overlap in BCCAs would be saturated.
This suggests that the relative error $\Delta$ would also become saturated with increasing $N$. 
In the case of $d_f=3$, as in BPCAs, the phase shift increases with increasing $N$; thus, the RGD theory is not applicable to BPCAs at sufficiently large $N$. 
If the phase shift due to the aggregates cannot be considered negligible, the forward scattered light rays are not in phase, and thus the forward scattered intensity might be slightly smaller than that of coherent scattering. As a result, the relative error in the forward scattering for the BPCA model increases with increasing $N$, as shown in Figure \ref{fig:error}.

\section{Concluding remarks} \label{sec:summary}
We investigated the angular dependence of scattering by two types of fractal dust aggregates, BCCAs and BPCAs, which have fractal dimensions of 2 and 3, respectively. 
Three methods were used to calculate the light scattering: the TMM, Mie theory with the EMT, and the RGD theory. 
In the formulation of the RGD theory, each phase matrix element of the fractal dust aggregates can be expressed as the product of the corresponding phase matrix element of the monomer and the structure factor. For the two-point correlation function of fractal aggregates, the Gaussian cut-off model was adopted (Equation (\ref{eq:gumodel}) with $\beta=2$). 
Employing this model, the structure factor is expressed analytically as a function of the aggregate radius $R_g$, the monomer radius $R_0$, and the fractal dimension $d_f$ (Equation (\ref{eq:gauss0}) or (\ref{eq:gauss})).  

Our results show that the RGD theory is a useful tool for calculating the phase function of fractal dust aggregates with $d_f\lesssim2$. The main conclusions of this study are summarized as follows.
\begin{itemize}
\item The RGD theory is consistent with both the wavelength and angle dependence of the scattered intensity (Section \ref{sec:S11}). The RGD theory agrees with the TMM with an accuracy of $\lesssim 10\%$ for the BCCA model with $N<1024$ (Section \ref{sec:relaerr}). 

 \item The angular distribution of scattered light can be understood to be dependent on the hierarchical structure of dust aggregates. 
The scattered intensity at small angles where all scattered light rays are in phase is determined by the large scale structure of the aggregates. 
In the case of large-angle scattering, the intensity depends on the degree of coherence of the scattered light; hence, the relative position of monomers comes into play. Therefore, the internal structure is responsible for the intensity at large scattering angles (Section \ref{sec:scalingap}). 

 \item The effective medium theory underestimates the backward scattering intensity by multiple orders of magnitude when $X_{\rm agg}>1$. 
 This is because the EMT ignores the internal structure of dust aggregates (Section \ref{sec:scalingap}).

\item Although the RGD theory tends to slightly overestimate the degree of linear polarization, it exhibits results that are qualitatively similar to the rigorous TMM calculation results (Section \ref{sec:LDP}). The occurrence of cross-polarization may be responsible for this overestimation (Appendix \ref{sec:crosspol}).
\end{itemize}

The applicability of the RGD theory to BCCAs is determined by whether the phase shift of the monomer is negligible.
In addition, for $X_m\gtrsim1$, monomer--monomer interactions are induced, which causes slight depolarization at $\theta=90^{\circ}$.
The quantitative modeling of polarization by fractal aggregates and phase matrix elements by opaque BCCA is a goal for future work, whereas transparent BCCA was investigated in this paper. 
In addition, the conditions of the RGD theory (Equations (\ref{eq:rgdcond1}), (\ref{eq:rgdcond2}), and (\ref{eq:rgdcond3})) should be tested over a wide range in parameter space. This is also a future objective.
Because of the large phase shift or the occurrence of multiple scattering, the light scattering properties of BPCAs containing a large number of monomers cannot be calculated using the RGD theory.

\acknowledgments
We appreciate useful comments by the referee, which helped us to improve the clarity of our discussion.
R.T. thanks Yasuhiko Okada for his technical advice regarding the T-matrix method and Takashi Kozasa for providing calculation data \citep{kozasa93}. R.T. would also like to acknowledge the valuable discussions he had with Hiroshi Kimura, Koji Murakawa, Munetake Momose, and Sanemichi Z. Takahashi. R.T. was supported by a Research Fellowship for Young Scientists from the Japan Society for the Promotion of Science (JSPS) (15J02840).
This work was supported by Grants-in-Aid for Scientific Research 23103005 and 25400229.

\appendix

\section{Rayleigh--Gans--Debye theory} \label{sec:appA}
Because electromagnetic waves are vector quantities, Equation (\ref{eq:RGD}) should be derived by vector analysis.
However, the nature of Equation (\ref{eq:RGD}) can be easily understood by analogy to the scalar wave scattering theory.
In this appendix, the derivation of Equation (\ref{eq:RGD}) is summarized by analogy to scalar wave scattering.

The propagation of a scalar wave $\psi$ in a medium obeys the Helmholtz equation, which is given by
\begin{equation}
\nabla^2{\psi}+k^2\psi=U(\vecbf{r})\psi, \label{eq:helmholtz}
\end{equation}
where $U(\vecbf{r})$ is the perturbing potential. If $U=0$, the solution $\psi$ to this equation is a plane wave. 
The solution to this equation can be written in the following form:
\begin{equation}
\psi(\vecbf{r})=\psi_0(\vecbf{r})+\int d\vecbf{r'} U(\vecbf{r'})G(\vecbf{r}-\vecbf{r'})\psi_{\rm inc}(\vecbf{r'}), \label{eq:scaintg}
\end{equation}
where $G(\vecbf{r})$ is an outgoing wave Greens function, $\psi_0$ is an incident plane wave, and $\psi_{\rm inc}$ is the sum of $\psi_0$ and multiple scattered light.
If the multiple scattering can be disregarded, $\psi_{\rm inc}\simeq \psi_0$ (first-order Born approximation).
Now, the scatterer is a collection of monomers, and the perturbing potential $U(\vecbf{r})$ can thus be expressed as the sum of the perturbing potentials of all monomers, $U(\vecbf{r})=\sum_{j=1}^{N} U_j(\vecbf{r})$, where $U_j(\vecbf{r})$ is the potential of the $j$th monomer. Assuming a spherical potential for all monomers yields 
\begin{equation}
U_j(r)=k^2(1-\epsilon_0)W(|\vecbf{r}-\vecbf{r_j}|,a), \label{eq:uj}
\end{equation}
where $\vecbf{r_j}$ the position vector of the $j$th monomer, $W(|\vecbf{r}-\vecbf{r'}|,a)$ is a window function in which $a$ represents the monomer radius, and $\epsilon_0$ is the dielectric function of monomer.
$W$ is unity when $|\vecbf{r}-\vecbf{r'}| \leq a$ and $0$ when $|\vecbf{r}-\vecbf{r'}|>a$. 

The second term of Equation (\ref{eq:scaintg}) has the form $f(\theta,\phi)e^{ikr}/r$ at large distances, and $f(\theta,\phi)$ is referred to as the scattering amplitude. Because a spherical symmetric potential is assumed, the scattering amplitude is reduced to $f(\theta)$. 
From Equations (\ref{eq:scaintg}) and (\ref{eq:uj}), the scattering amplitude of the aggregates can be described as 
\begin{equation}
f(\theta)=-\frac{1}{4\pi}k^2(1-\epsilon_0)\int d\vecbf{r'} \sum_{j=1}^{N}W(|\vecbf{r'}-\vecbf{r_j}|,a)e^{-i({\scriptsize \vecbf{k_s}-\vecbf{k_i}})\cdot{\scriptsize {\vecbf{r'}}}}.
\end{equation}
By substituting $\vecbf{R}=\vecbf{r'}-\vecbf{r_j}$ and $\vecbf{q}=\vecbf{k_s}-\vecbf{k_i}$, the variables can be separated as
\begin{equation}
f(\theta)=-\frac{1}{4\pi}k^2(1-\epsilon_0)\int d\vecbf{R} W(R,a)e^{-i{\scriptsize \vecbf{q}}\cdot {\scriptsize {\vecbf{R}}}}\sum_{j=1}^{N}e^{-i{\scriptsize \vecbf{q}}\cdot{\scriptsize {\vecbf{r_j}}}},
\end{equation}
where $R=|\vecbf{R}|$. Because the differential scattering cross section $dC_{\rm sca}/d\Omega$ equals the square of the scattering amplitude,
\begin{equation}
\frac{dC_{\rm sca}}{d\Omega}=\frac{1}{16\pi^2}k^4|\epsilon-1|^2v^2\left|\frac{1}{v}\int d\vecbf{R} W(R,a)e^{-i{\scriptsize \vecbf{q}}\cdot {\scriptsize {\vecbf{R}}}}\right|^2
\left|\sum_{j=1}^{N}e^{-i{\scriptsize \vecbf{q}}\cdot{\scriptsize {\vecbf{r_j}}}} \right|^2, \label{eq:diffscagen}
\end{equation}
where $v=4\pi{a}^3/3$ is the volume of the monomer.
The Gans form factor is defined as \citep{gans1925, bohren83}
\begin{equation}
F(\vecbf{q})\equiv\frac{1}{v}\int d\vecbf{R} W(R,a)e^{-i {\scriptsize \vecbf{q}}\cdot{{\scriptsize \vecbf{R}}}}. \label{eq:gansF}
\end{equation}
From Equations (\ref{eq:diffscagen}) and (\ref{eq:gansF}), the following relationship can be obtained:
\begin{equation}
\frac{dC_{\rm sca}}{d\Omega}=\frac{1}{9}k^4a^6|\epsilon-1|^2F(\vecbf{q})^2\left|\sum_{j=1}^{N}e^{-i{\scriptsize \vecbf{q}}\cdot{\scriptsize {\vecbf{r_j}}}} \right|^2.
\end{equation}
The scattering amplitude for the monomer can be described as $f_{l}=\sin\delta_l\exp(i\delta_l)/k$, where $\delta_l$ is the phase shift of an $l$-wave. Assuming $ka\ll1$, the phase shift of the lowest order of a partial wave ($l=0$ or s-wave) becomes approximately $\delta_0\simeq\frac{1}{3}(ka)^3|\epsilon-1|$ \citep[e.g.,][]{Landau65}. Thus,
\begin{equation}
\frac{dC_{\rm sca}}{d\Omega}=\left[\frac{dC_{\rm sca}}{d\Omega}\right]_{\rm Rayleigh}F(\vecbf{q})^2\left|\sum_{j=1}^{N}e^{-i{\scriptsize \vecbf{q}}\cdot{\scriptsize {\vecbf{r_j}}}} \right|^2, \label{eq:diffsca2}
\end{equation}
where $[dC_{\rm sca}/d\Omega]_{\rm Rayleigh}=|f_0|^2$.
The last summation term is called the Debye factor. 
Using Equations (\ref{eq:cordef}) and (\ref{eq:nnorm}), we obtain
\begin{equation}
N^{-2}\left|\sum_{j=1}^{N}e^{-i{\scriptsize \vecbf{q}}\cdot{\scriptsize {\vecbf{r_j}}}}\right|^2=\int g(\vecbf{u})e^{i {\scriptsize \vecbf{q}}\cdot{{\scriptsize \vecbf{u}}}}d\vecbf{u}, \label{eq:debye}
\end{equation}
Using Equations (\ref{eq:diffsca2}), (\ref{eq:debye}), and (\ref{eq:wk}) yields
\begin{equation}
\frac{dC_{\rm sca}}{d\Omega}=N^2\left[\frac{dC_{\rm sca}}{d\Omega}\right]_{\rm Rayleigh}F(\vecbf{q})^2
\mathcal{S}(\vecbf{q}).
\end{equation}
Thus, the differential cross section of the aggregates can be given in terms of the differential Rayleigh cross section; the Gans form factor; and the Debye, or structure, factor. In this regard, this approach is called the Rayleigh--Gans--Debye theory.
Replacing the Rayleigh--Gans term with the exact Mie solution for a spherical monomer and using $S_{11}/k^2=dC_{\rm sca}/d\Omega$ yields
\begin{equation}
S_{11, {\rm agg}}=N^2S_{11, {\rm mono}}\mathcal{S}(\vecbf{q}).
\end{equation}

\section{Confluent hypergeometric function}  \label{sec:appB}
\begin{figure}[t]
\begin{center}
\includegraphics[height=7.0cm,keepaspectratio]{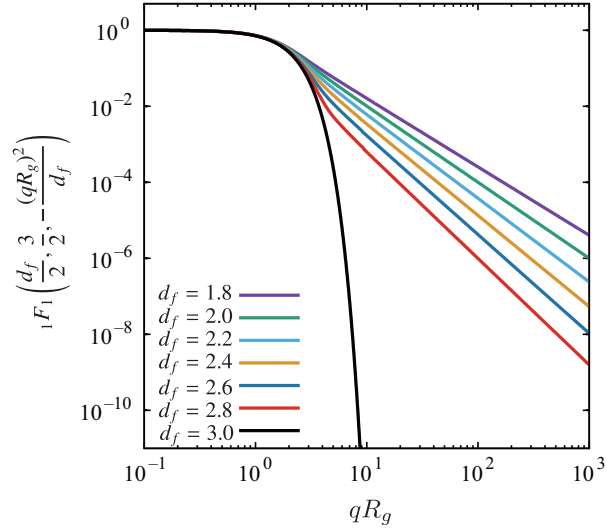}
\caption
{Confluent hypergeometric function for various fractal dimensions $d_f$.}
\label{fig:1F1}
\end{center}
\end{figure}

The confluent hypergeometric function is defined as
\begin{equation}
{}_1F_1(\alpha;\beta;z)=\sum_{n=0}^{\infty} \frac{(\alpha)_{n}z^n}{(\beta)_{n}n!} \label{eq:1F1},
\end{equation}
where $(x)_n$ ($x=\alpha, \beta$) is the Pochhammer symbol given by
\begin{eqnarray}
(x)_0&=&1\\
(x)_n&=&\prod_{k=0}^{n-1}(x+k).
\end{eqnarray}
If $\alpha=\beta$, ${}_1F_1(\alpha;\beta;z)=\sum_{n=0}^{\infty} z^n/n!=\exp(z)$.
It is quite useful to use the asymptotic form of confluent hypergeometric function for $|z|\gg1$; hence,
\begin{equation}
{}_1F_1(\alpha;\beta;z)\simeq\frac{\Gamma(\beta)}{\Gamma(\beta-\alpha)}(-z)^{-\alpha}\sum_{n=0}^{\infty}(-1)^n\frac{(\alpha)_n(\beta-\alpha)_n}{n!z^n}+\frac{\Gamma(\beta)}{\Gamma(\alpha)}e^zz^{\alpha-\beta}\sum_{n=0}^{\infty}\frac{(1-\alpha)_n(\beta-\alpha)_n}{n!z^n}. \label{eq:1f1asym}
\end{equation}
When $z$ has a large negative value, Equation (\ref{eq:1f1asym}) reduces to
\begin{equation}
{}_1F_1(\alpha;\beta;-|z|)\simeq\frac{\Gamma(\beta)}{\Gamma(\beta-\alpha)}|z|^{-\alpha}\sum_{n=0}^{\infty}\frac{(\alpha)_n(\beta-\alpha)_n}{n!|z|^n}. \label{eq:1f1asym2}
\end{equation}
Equation (\ref{eq:1f1asym2}) shows that for $z\gg1$, the summation decreases to approximately unity so that the asymptotic form of the confluent hypergeometric function gives rise to a simple power law function. Substituting $\alpha=d_f/2$, $\beta=3/2$, and $|z|=(qR_g)^2/d_f$ yields
\begin{eqnarray}
{}_1F_1\left(\frac{d_f}{2},\frac{3}{2},-\frac{(qR_g)^2}{d_f}\right)&\simeq& C(d_f)(qR_g)^{-d_f},\\
C(d_f)&=&\frac{\sqrt{\pi}}{2}\frac{d_f^{d_f/2}}{\Gamma(\frac{3-d_f}{2})}.
\end{eqnarray}
Consequently, $d_f=2$ gives $C=1$ because $\Gamma(1/2)=\sqrt{\pi}$. Equation (\ref{eq:1F1}) with $\alpha=d_f/2$, $\beta=3/2$, and $z=-\frac{(qR_g)^2}{d_f}$ is plotted for various $d_f$ in Figure \ref{fig:1F1}.

\section{Depolarization effect of aggregates} \label{sec:crosspol}
\begin{figure*}[t]
\begin{center}
\includegraphics[height=6.5cm,keepaspectratio]{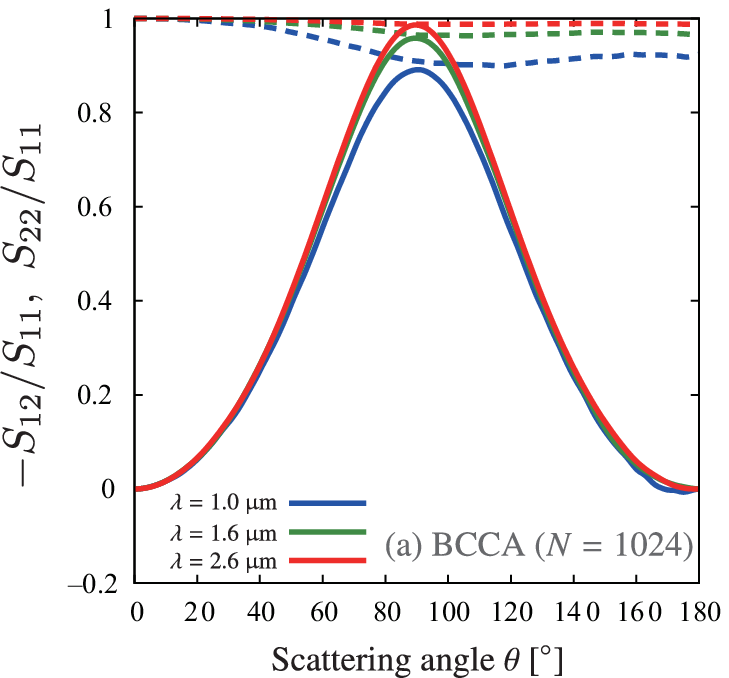}
\includegraphics[height=6.5cm,keepaspectratio]{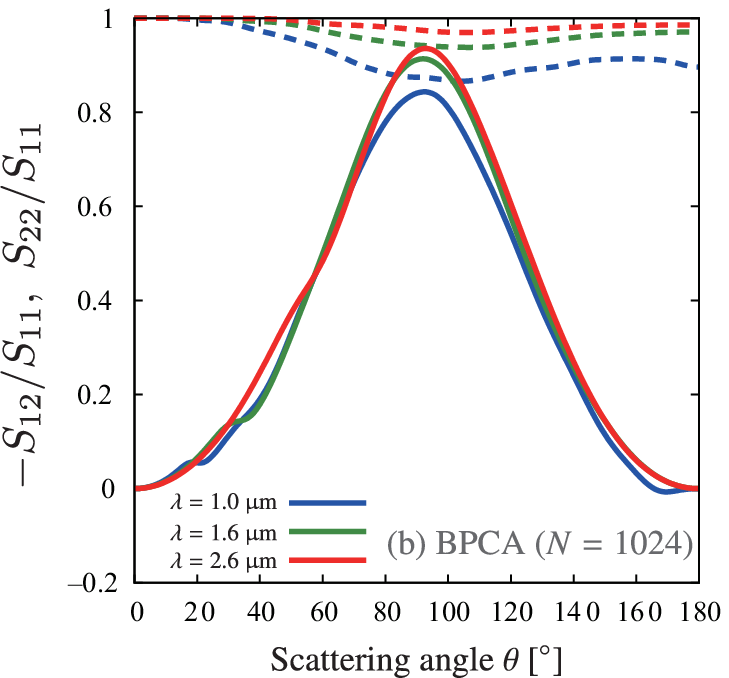}
\caption
{Degree of linear polarization $P$ and $S_{22}/S_{11}$ for (a) BCCA and (b) BPCA models with $N=1024$. The red, green, and blue lines represent $\lambda=2.6$, $1.6$ and $1.0\ \upmu$m. 
The solid lines indicate the linear degree of polarizability $P=-S_{12}/S_{11}$, and the dashed lines represent the ratio $S_{22}/S_{11}$ obtained using the TMM.}
\label{fig:1024pol}
\end{center}
\end{figure*}

As discussed in Section \ref{sec:LDP}, despite the fact that monomers are Rayleigh scatterers, the degree of linear polarization from the aggregates is slightly reduced at $\theta=90^{\circ}$. This appendix discusses the depolarization effect of aggregates. The depolarization is found to be caused by the occurrence of cross-polarization, which may increase $S_{11}$ and decrease $S_{12}$, thereby reducing the degree of polarization.
In this regard, the depolarization of dust aggregates is essentially different from the case of a single sphere in which cross-polarization does not occur.
When cross-polarization occurs, scattered light has a component that is perpendicular to the scattering plane even if the incident light only has a parallel component, and vice versa. In other words, $S_3$ and $S_4$ of the scattering amplitude matrix elements are not zero (see Chap. 3 of BH83). The occurrence of cross-polarization can be determined based on the ratio $S_{22}/S_{11}$ because $S_{22}/S_{11}$ is less than unity whenever cross-polarization occurs.
Figure \ref{fig:1024pol} shows the degree of linear polarization $P$ and $S_{22}/S_{11}$ for BCCA and BPCA models with $N=1024$.
Figure \ref{fig:1024pol} demonstrates that $S_{22}/S_{11}$ equals unity at small scattering angles, whereas it is less than unity at large scattering angles. The reason $S_{22}/S_{11}=1$ at small scattering angles is because the aggregate can be regarded as a single sphere owing to the coherent scattering (see Section \ref{sec:scalingap}).
The maximum value of the degree of linear polarization correlates with $S_{22}/S_{11}$. Therefore, to determine the maximum degree of polarization of the aggregates, cross-polarization should be considered, which is not the case in the RGD theory.
	
Next, the origin of the cross-polarization is discussed. A possible mechanism of this depolarization is the monomer--monomer interaction.
Because Rayleigh scattering shows completely polarized scattered light at $\theta=90^{\circ}$, the light scattered by the aggregates also shows completely polarized light as long as the interaction between monomers is disregarded. 
Therefore, the depolarization can be interpreted as a consequence of monomer--monomer dipole interactions \citep{lu94}.
\citet{berry86} argued the importance of the multiple scattering of fractal aggregates by means of the mean field approximation and concluded that multiple scattering can be negligible for fractal aggregates of small monomers with $d_f\leq2$, like BCCAs. However, even if $d_f\leq2$, monomer--monomer interactions cannot be considered negligible for large monomers. 
\citet{okada09} and \citet{mishchenko13} found that the linear depolarization ratio can be used as a diagnosing tool for the density of the aggregates. 
A monomer in a dense aggregate, like a BPCA, tends to interact with many nearby monomers; therefore, it is expected that the depolarization effect is more prominent for dense aggregates than for fluffy aggregates.
Indeed, Figure \ref{fig:1024pol} shows that BPCAs tend to show more depolarized scattering than BCCAs.

Because of the monomer--monomer interaction, the polarized vector excited by each monomer is not always parallel to the external incident field. Hence, these interacting monomers might be treated approximately as anisotropic Rayleigh spheres with different polarizabilities with respect to the three different axes.
\citet{mishchenko13} investigated the modeling of linear depolarization by aggregates using the formula of an anisotropic Rayleigh sphere:
\begin{equation}
P=\frac{1-\cos^2\theta}{y+\cos^2\theta},
\end{equation}
where $y$ is the anisotropy parameter, and it varies from $1$ to $13$ (see Equations (5.53) and (5.54) of BH83). In the case of an isotropic sphere, $y=1$ so that the polarizability in $\theta=90^{\circ}$ is $100\%$. Note that the anisotropic sphere shows a symmetric profile with respect to $\theta=90^{\circ}$, whereas aggregates show slightly asymmetric profiles. Our results for BCCA and BPCA models with $\lambda=1\ \upmu$m are $y\simeq1.1$ and $y\simeq1.2$, respectively.
To model the depolarization of the fractal aggregates, it is important to investigate how $y$ varies as a function of the number $N$ of monomers, the monomer radius $R_0$, the aggregate composition, and the fractal dimension $d_f$. This remains as an objective for a future study.

\bibliographystyle{apj}

\bibliography{RTetal2016}

\end{document}